\documentclass[letterpaper,twocolumn,10pt]{article}
\usepackage{usenix-2020-09}

\usepackage{enumitem}  
\usepackage[utf8]{inputenc}
\usepackage[T1]{fontenc}  
\usepackage{booktabs}
\usepackage{tikz}
\usepackage{xspace}
\usepackage{amsthm}
\usepackage{latexsym}
\usepackage{multirow}
\usepackage{makecell}
\usepackage[normalem]{ulem}
\usepackage{amsmath}
\usepackage{amsfonts}
\usepackage{pifont}
\usepackage{xcolor}
\usepackage{hyperref}
\usepackage{cleveref}
\usepackage{url}
\usepackage{soul}
\usepackage{algorithmic}
\usepackage{algorithm}
\usepackage{makecell}
\usepackage{colortbl}
\usepackage{caption}
\usepackage{subcaption}
\usepackage{calc}  
\usepackage[nomessages]{fp}
\usepackage[autolanguage]{numprint}

\newcommand{\cmark}{\ding{51}}%
\newcommand{\xmark}{\ding{55}}%

\usepackage[url=true,isbn=false,doi=false,style=trad-plain,maxbibnames=99,sortcites]{biblatex}
\AtBeginBibliography{\small}
\AtEveryBibitem{%
  \clearfield{note}%
  \clearfield{series}%
}
\AtEveryBibitem{\clearfield{month}}
\AtEveryBibitem{\clearfield{day}}
\DeclareSourcemap{
  \maps[datatype=bibtex, overwrite]{
    \map{
      \step[fieldset=pages, null]
      \step[fieldset=language, null]
      \step[fieldset=publisher, null]
      \step[fieldset=eventtitle, null]
      \step[fieldset=address, null]
      \step[fieldset=location, null]
    }
  }
}

\newtheorem{thm}{Theorem}[section] 
\theoremstyle{plain} 
\newcommand{\thistheoremname}{}
\newtheorem{genericthm}[thm]{\thistheoremname}

\definecolor{orange}{rgb}{1,0.5,0}

\newcommand{\lyx}[1]{{\color{black}{#1}}}

\newcommand{\etal}{\textit{et al.}\xspace}

\newcommand{\remove}[1]{}

\newcommand{\sys}{\texttt{ConsentChk}\xspace}

\newcommand{\devtools}{DevTools\xspace}

\newcommand{\cookiebot}{Cookiebot\xspace}
\newcommand{\onetrust}{OneTrust\xspace}

\newcommand{\numSitesAvailStudy}{10k\xspace}

\FPeval{\numSucSites}{9159}

\newcommand{\topThreeScore}{85.96\%\xspace}

\FPeval{\numDomainsToLabel}{1000}
\FPeval{\numLabeledWebsites}{1000}
\FPeval{\numWebsitesWithSetting}{192}
\FPeval\numWebsitesWithSettingPercent{round(\numWebsitesWithSetting/\numLabeledWebsites*100,1)}
\FPeval{\numWebsitesNoSetting}{clip(\numLabeledWebsites - \numWebsitesWithSetting)}

\FPeval{\numSitesWithSettingButNoBanner}{68}
\FPeval{\numSitesWithSettingButNoBannerPercent}{round(\numSitesWithSettingButNoBanner / \numWebsitesWithSetting * 100, 1)}
\FPeval{\numBannersHideSetting}{3}
\FPeval{\numBannersHideSettingPercent}{round(\numBannersHideSetting / \numWebsitesWithSetting * 100, 1)}
\FPeval{\numBannersWithSetting}{121}
\FPeval{\numBannersWithSettingPercent}{round(\numBannersWithSetting / \numWebsitesWithSetting * 100, 1)}

\FPeval{\numSettingButNoBanners}{clip(\numWebsitesWithSetting - \numBannersWithSetting)}
\FPeval{\numSettingButNoBannersPercent}{round(\numSettingButNoBanners / \numWebsitesWithSetting * 100, 1)}

\FPeval{\numBannersBinaryChoice}{15}
\FPeval{\numBannersBinaryChoicePercent}{round(\numBannersBinaryChoice / \numWebsitesNoSetting * 100, 1)}

\FPeval{\numBannersInNoSettingSites}{135}
\FPeval{\numBannersInNoSettingSitesPercent}{round(\numBannersInNoSettingSites / \numWebsitesNoSetting * 100, 1)}

\FPeval{\numThirdPartyCookieLibs}{19}
\FPeval{\firstPartyCookieLibCount}{57}
\FPeval{\firstPartyCookieLibCountPercent}{round(\firstPartyCookieLibCount / \numWebsitesWithSetting * 100, 1)}

\FPeval{\thirdPartyCookieLibCountPercent}{round(100 - \firstPartyCookieLibCountPercent, 1)}
\FPeval{\numThirdPartySiteCount}{clip(\numWebsitesWithSetting - \firstPartyCookieLibCount)}

\FPeval{\numOneTrustSites}{66}
\FPeval{\numOneTrustSitesPercent}{round(\numOneTrustSites / \numWebsitesWithSetting*100, 1)}

\FPeval{\numTcfWebsites}{63}
\FPeval{\numTcfWebsitesPercent}{round(\numTcfWebsites / \numLabeledWebsites * 100, 1)}
\FPeval{\numTcfWithCookset}{27}
\FPeval{\numTcfWithCooksetPercent}{round(\numTcfWithCookset / \numWebsitesWithSetting * 100, 1)}
\FPeval{\numTcfWithNoCookset}{clip(\numTcfWebsites - \numTcfWithCookset)}
\FPeval{\numTcfSitesTopTenK}{681}
\FPeval{\numTcfSitesTopTenKPercent}{round(\numTcfSitesTopTenK / \numSucSites * 100, 1)}

\newcommand{\evalNumSitesCookieSettingStudy}{20k\xspace}
\FPeval{\nScannedSites}{10436}

\FPeval{\nSitesWithSettings}{1458} 
\FPeval{\nSitesWithSettingsPercent}{round(\nSitesWithSettings / \nScannedSites*100, 2)} 
\FPeval{\nSitesWithSettingsUk}{1433}
\FPeval{\nSitesWithSettingsCa}{1401}
\FPeval{\nSitesCaLessUkPercent}{round((\nSitesWithSettingsUk - \nSitesWithSettingsCa) / \nSitesWithSettingsUk*100, 2)}
\FPeval{\nSitesCaLessEuPercent}{round((\nSitesWithSettings - \nSitesWithSettingsCa) / \nSitesWithSettings*100, 2)}

\FPeval{\evalWebsiteWithOneTrust}{1340}
\FPeval{\evalWebsiteWithCookiebot}{110}
\FPeval{\evalWebsiteWithTermly}{clip(\nSitesWithSettings - \evalWebsiteWithOneTrust - \evalWebsiteWithCookiebot)}

\FPeval{\nsiteContra}{154}

\FPeval{\nsiteOmit}{1291}
\FPeval{\nsiteIncor}{1204}
\FPeval{\nsiteIncorPercent}{round(\nsiteIncor / \nSitesWithSettings*100, 2)} 
\FPeval{\nsiteAmbi}{66}
\FPeval{\nsiteComply}{71}

\FPeval{\nsiteDura}{1025}
\FPeval{\nsiteDuraPercent}{round(\nsiteDura / \nSitesWithSettings*100, 2)}

\FPeval{\nsiteExtraUn}{143}
\FPeval{\nsiteExtraUnPercent}{round(\nsiteExtraUn / \nSitesWithSettings*100, 2)}

\FPeval{\nsiteNarrow}{147}
\FPeval{\nsiteNarrowPercent}{round(\nsiteNarrow / \nSitesWithSettings*100, 2)}

\FPeval{\nsitePreaccept}{672}
\FPeval{\nsitePreacceptPercent}{round(\nsitePreaccept / \nSitesWithSettings*100, 2)}

\FPeval{\nsitePrefbtn}{185}

\newcommand{\timeoutSec}{60}

\newcommand*\nt[2][z]{%
  \ifx z#1%
    $\numprint{#2}$%
  \else%
    $\numprint[#1]{#2}$%
  \fi%
  \xspace
}

\newcommand*\pp[3][v]{%
  \ifx q#1%
    \nt{#2}/\nt{#3}(\ShowPercentage{#2}{#3})%
  \else%
    \ifx p#1%
      \nt{#2} (\ShowPercentage{#2}{#3})\else%
    \ifx c#1%
      \ShowPercentage{#2}{#3}(\nt{#2}/\nt{#3})\else%
    \ifx r#1%
      \ShowPercentage{#2}{#3}\else%
    \nt{#2}%
  \fi
\fi\fi\fi
}

\newcommand*\np[2][z]{
  \ifx z#1
    \nprounddigits{2}$\numprint{#2}$
  \else
    \nprounddigits{2}$\numprint[#1]{#2}$
  \fi
}

\newcommand{\ShowPercentage}[2]{\FPeval\percentage{round(#1/#2*100,0)}\FPeval\percentageOneDecimal{round(#1/#2*100,2)}\percentageOneDecimal\%\xspace}

\begin{document}

\pagestyle{plain}
\title{Navigating Cookie Consent Violations Across the Globe}


\author{
\thanks{These authors contributed equally to this work}
{\rm Brian Tang\footnotemark[1]}\\
University of Michigan\\
{\rm \texttt{bjaytang@umich.edu}}
\and
{\rm Duc Bui\footnotemark[1]}\\
University of Michigan\\
{\rm \texttt{ducbui@umich.edu}}
\and
{\rm Kang G. Shin}\\
University of Michigan\\
{\rm \texttt{kgshin@umich.edu}}
}











\maketitle     
\begin{abstract}

Online services provide users with 
{\em cookie banners} to accept/reject 
the cookies placed on their web browsers.
Despite the increased adoption of cookie banners, 
little has been done to ensure that cookie consent is compliant with privacy laws around the globe. 
Prior studies have found that cookies are often placed 
on browsers even after their explicit rejection by users.
These inconsistencies in cookie banner behavior circumvent
users' consent preferences and are
known as \textit{cookie consent violations}.
To address this important problem, we propose an 
end-to-end system, called \sys, that 
detects and analyzes cookie banner behavior. 
\sys uses a formal model to systematically 
detect and categorize cookie consent violations.
We investigate eight English-speaking regions 
across the world, and analyze cookie banner behavior 
across 1,793 globally-popular websites. 
Cookie behavior, cookie consent violation rates, and 
cookie banner implementations are found to be highly 
dependent on region. 
Our evaluation reveals that consent management 
platforms (CMPs) and website developers likely tailor 
cookie banner configurations based on their (often 
incorrect) interpretations of regional privacy laws.
We discuss various root causes behind these 
cookie consent violations. 
The resulting implementations produce misleading 
cookie banners, indicating the prevalence 
of \textit{inconsistently implemented and enforced}
cookie consent between various regions.
\end{abstract}



\section{Introduction}

\begin{figure}[t]
    \centering
    \includegraphics[width=0.9\columnwidth]{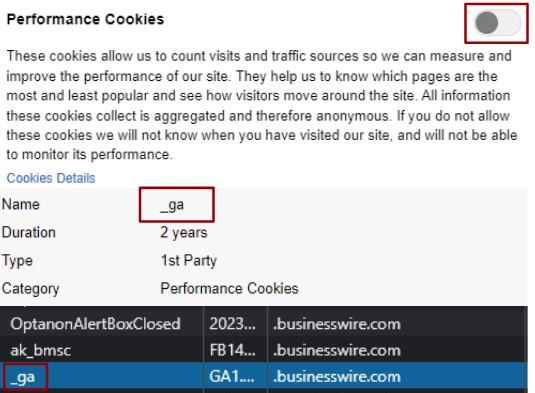}
    \caption{
    A cookie banner menu allows users to set their 
    consent/rejection of cookies.
    However, the website fails to honor the users' consent (e.g., Google Analytics).
    }
    \label{fig:cookie_setting_example}
\vspace*{-0.15in}
\end{figure}

One of the most common ways of collecting data from 
users' web browsers is through the use of cookies. 
Cookies are used to remember users' preferences, 
transmit information such as location or IP addresses, and 
track users' browsing history. However, the collection and use 
of personal data through cookies have raised significant 
privacy concerns. In response to these concerns, regions/countries 
worldwide have enacted laws requiring disclosure of 
privacy practices and/or user consent for data collection. 
For instance, the EU's General Data 
Protection Regulation (GDPR)~\cite{gdpr_2016} and ePrivacy 
Directive (ePD)~\cite{epd_2002} suggest online services 
obtain consent before collecting personal data for advertising. 
Similarly, the California Consumer Privacy Act (CCPA) in the 
U.S., Canada's Personal Information Protection and Electronic 
Documents Act (PIPEDA), and Singapore's Personal Data 
Protection Act (PDPA), are among other similar privacy 
regulations that provide guidelines on consent, online behavioral 
advertising (OBA), and tracking cookies. These policies reflect 
a global shift towards enhanced data privacy protections.

To comply with these regulations, websites employ mechanisms and 
third-party services known as \textit{Consent Management Platforms} 
(CMPs) to allow users to manage their cookie preferences. 
Figure.~\ref{fig:cookie_setting_example} illustrates an example 
of rejecting cookies from Google Analytics. 
Despite these measures, prior research indicates that cookies are 
often placed on browsers even after users have explicitly rejected 
them through cookie settings 
interfaces~\cite{bollinger2022automating,matte_cookie_2020}.
These inconsistencies in cookie banner behavior circumvent users'
consent preferences and are
known as \textit{cookie consent violations} or cookie violations.
Such practices not only undermine user trust but may also go 
against regional privacy 
guidelines~\cite{federal_trade_commission_goldenshores_2013,
european_data_protection_board_belgian_2019,dataguidance_france_2021}.

These observations prompt three critical questions:
\vspace*{-0.04in}
\begin{enumerate}[noitemsep,leftmargin=0.4cm,topsep=5pt]
  \item \textit{How pervasive are cookie consent violations?}
  \item \textit{Do cookie banners, behaviors, and violations vary 
    across different regions with distinct privacy regulations?}
  \item \textit{Why do these discrepancies in cookie behaviors occur?}
\end{enumerate}

We address these questions by conducting a comprehensive 
cross-country analysis of cookie consent practices.
We examine websites 
across 8 English-speaking regions with varying privacy regulations---Ireland (EU), the United Kingdom (UK), 
California, USA (CA), Michigan, USA (US), Canada (CAN), 
South Africa (ZA), Singapore (SG), and Australia (AU)---to understand 
how cookie consent mechanisms are implemented and whether they are 
consistent with users' choices and regional privacy laws. 
We have designed and deployed a cookie consent measurement and auditing 
tool called \sys that automatically interacts with cookie banners.
We use this tool to analyze cookie consent and CMPs on websites 
across these 8 regions worldwide.

\sys addresses several technical challenges 
in detecting cookies that violate user consent. 
It detects and activates cookie banners' menus, 
setting the consent choice for each cookie (accept/reject), 
and checks for inconsistencies in the expected and observed 
behavior. We conduct an evaluation analyzing cookie consent 
behavior for the same set of websites across different regions 
with different regulatory privacy frameworks. 
Using \sys, we have examined 1,793 websites from the 
\lyx{\evalNumSitesCookieSettingStudy} 
top global websites according to the Tranco list
\cite{LePochat_tranco_2019}.

From these measurements, we found widespread 
occurrences of cookie consent violations across all 8 regions. 
Websites exhibited higher rates of violations 
in certain regions than others. 
Our analyses indicate that cookie consent is
\textit{inconsistently implemented and enforced} in
all regions. By studying the occurrence 
of cookie consent violations in these regions,
we uncover several potential root causes. In particular, due to how
cookie consent libraries are implemented, the number of cookies and
violations is highly dependent on location. The ability
to reject consent for a particular cookie depends on whether
a CMP decides to support privacy laws via a template.





This paper makes the following contributions:
\vspace*{-0.04in}
\begin{itemize}[noitemsep,leftmargin=0.4cm,topsep=5pt]
  \item An end-to-end measurement tool, called \sys, that crawls 
    websites, detecting infringements on users' consent preferences 
    for each cookie on a website. \sys uses an approach 
    that identifies features of cookie banner 
    buttons to activate the cookie banner menus of 
    any website automatically. The detection 
    of cookie consent violations has a high precision of 
    \lyx{>91\%}. (\Cref{sec:design,sec:evaluation})
  \item A study on 1,793 of the top visited 
    websites in the globe based on the Tranco 1M 
    list~\cite{LePochat_tranco_2019} with cookie banners. 
    \sys detected that 96.18\% (EU) -- 97.72\% (US) of websites across all 
    regions contain at least one cookie consent violation. 
    These violations primarily occur in the form of undeclared 
    cookies or ignored cookie rejections, demonstrating
    that cookie banners and CMPs have inadequate coverage and correctness. (\Cref{sec:methodology,sec:study})
  \item The discovery of discrepancies in cookie consent
    behavior across various regions. We observe that, compared to 
    the EU, the same websites in the US have an average of almost 
    12 additional cookies, and an average of 10 additional cookie 
    consent violations. 
    Similarly, regions like Canada, Singapore, South Africa, and 
    Australia with their own privacy frameworks have 6-8 more 
    cookies and 5-7 more cookie consent violations than the EU.
    (\Cref{sec:study})
  \item An analysis of the potential root causes behind cookie 
    consent violations and regional discrepancies. CMPs provide banner customization and geolocation rulesets. Thus, cookie banner buttons and cookie behaviors are largely 
    reliant on factors such as location, advertiser interest, 
    cookie scraping/classification accuracy, and more. 
    (\Cref{sec:root_cause})
\end{itemize}



\section{Background}\label{sec:background}

\begin{table*}[t]
    \footnotesize
    \centering
    \begin{tabular}{l|l|l|l|l|l}
    \toprule
    \textbf{Region} & \textbf{Law} & \textbf{Relevant Articles and Sections} & \textbf{E.T.} & \textbf{Consent} & \textbf{Personal Information Definition} \\
    \midrule
    \midrule
    EU & GDPR & Articles 3, 4, 7, 21, 82.1, Recital 30, 32 & Yes & Informed & Info About Person, Includes Cookies, Identifiers\\
    \midrule
    UK & DPA & Section 57, 99 & Yes & Informed & Info About Person, Includes Cookies, Identifiers \\
    \midrule
    CA (US) & CCPA & Section 1798.120, 1798.140.aj, 1798.145.a1G, 7026.h & Yes & Implied & Any Unique ID, Includes Cookies, Identifiers \\
    \midrule
    MI (US) & None & None & None & None & None  \\
    \midrule
    CAN & PIPEDA & Schedule 1, Principle 3 (4.3.8), Sec 1.6.1 & Yes~\cite{canada2017courtruling} & Informed & Info About Person, Includes Cookies, Identifiers \\
    \midrule
    ZA & POPIA & Sections  3.1(b), 5(d)-(e), and 11(3)-(4), Ch. 1 & No & Informed & Info About Person, Identifiers \\
    \midrule
    SG & PDPA & Part 4 Division 1.16, Part 6 Division 26.1 & Yes & Implied & Identifiers, Includes Cookies/OBA \\
    \midrule
    AU & APP & APP 5.1, 6, 7 & No & Required & Identifiers \\
    \bottomrule
    \end{tabular}
    \caption{A brief high-level overview of region-specific privacy frameworks and their requirements. The E.T. column indicates whether there is an extra-territorial scope. Identifiers include IP, location, targeting cookies, biometrics, derivable identity, etc.} \label{tab:frameworks}
\end{table*}

\subsection{Privacy Laws Across the Globe}

Privacy laws generally provide rules for how data 
controllers/processors and businesses should process data. These laws also 
establish and define the following criterion required for: 
(1) establishing consent of data processing given by the subject, 
(2) defining what types of data fall under this protection 
(usually personal data, identifiers, and/or cookies). Several privacy laws like the GDPR and CCPA are also applied in an extraterritorial manner to Europeans and Californians who use non-European and non-Californian services.

{\bf GDPR (European Union, EU).}
Consent is one of 6 lawful bases defined in the GDPR for processing 
data. The others include a contractual obligation, a legal obligation, 
vital interests, a task in public interest, and legitimate interests. 
Many websites provide a privacy policy or notice for the data they 
process under these other 5 bases (for purposes such as marketing, 
fraud prevention, etc.), but additionally provide a cookie banner 
for obtaining consent for the usage of cookies.
Article 7 of the General Data Protection Regulation (GDPR) 
discusses informed consent related to the processing of 
users' personal data. In particular,
Article 7.3 states that data subjects can withdraw their 
consent at any time and that it should be as easy to 
withdraw as to give consent~\cite{eu:gdpr}.

{\bf DPA (United Kingdom, UK).}
The UK's Data Protection Act (DPA) is similar to GDPR,
following and adapting the same protection rules to the 
UK's legal system \cite{cookiebot_new_2021,
information_commissioners_office_uk_2022,part2018dpa}. 

{\bf CCPA (California, CA).}
Some privacy laws, such as the California Consumer Privacy Act (CCPA) 
do not require consent for data processing. Unlike the GDPR,
a lawful basis for processing data is not required, however, 
businesses cannot process data in unfair or deceptive manners. 
Instead, this framework requires that businesses meeting certain 
criteria (large amount of revenue from selling personal information, 
with certain exemptions) give consumers information about data 
collection activities and purposes in a ``notice at 
collection''~\cite{ccpa2023}, giving users the right to access, 
delete, and opt out of the selling of their data.

{\bf PIPEDA (Canada, CAN).}
Canada's Personal Information Protection and Electronic 
Documents Act (PIPEDA's) applies to organizations that collect, use, 
or disclose personal information in the course of a commercial 
activity. Fair Information Principle 3 states organizations must 
generally obtain express consent when the information being 
collected, used, or disclosed is sensitive~\cite{pipeda2025,pipeda2021tracking}. 

{\bf POPIA (South Africa, ZA).}
Similarly to the GDPR, the Protection of Personal Information 
Act (POPIA) identifies 5 legal bases for processing data, 
including consent from the data subject. It requires
organizations to obtain consent 
before collecting and processing personal information. Additionally, data subjects have the right to be notified of data 
collection, request correction, destruction, or deletion, and to 
object to processing of data for marketing~\cite{popia2019,popia2019consent}.

{\bf PDPA (Singapore, SG).}
The Personal Data Protection Act (PDPA) states in 
Part 4, Division 1 that an organization must not use 
or disclose personal data about an individual unless 
the individual gives, or is deemed to have given, his/her  
consent to the collection, use, or disclosure.
It also states that users may at any time withdraw 
any consent given in respect of the collection, use, 
or disclosure by that organization of personal data 
about the individual~\cite{pdpa2012}.

{\bf APP (Australia, AU).}
Section 5.1 of the Australian Privacy Principles (APP) 
requires entities to disclose purposes and request consent 
at or before the time or, if that is not practicable, 
as soon as practicable after, collecting personal 
information about an individual.
Section 6 requires entities to not use or 
disclose the collected personal information for another 
purpose without consent~\cite{app2025,app2024cookies}.

\subsection{Role of Cookie Consent in Privacy Laws}\label{subsec:legal}


While consent is one of many frameworks for allowing users to 
control and understand privacy practices, it is the primary form of 
control used for cookies on websites, due to the prevalence of cookies in online behavioral advertising. In Europe, the ePrivacy 
Directive (ePD) was an EU directive which modernized privacy laws to 
apply to cookies~\cite{epd_2002}. Cookies (other than those used for 
strictly necessary services) required user consent. These new 
guidelines paved the way for the IAB's Transparency and Consent 
Framework (TCF) to be created~\cite{iab_europe_tcf_2021}. The TCF is 
a voluntary standard providing publishers (websites) and advertisers 
(vendors) with tools and libraries for providing and managing 
consent options for users.

\begin{table*}[ht]
\footnotesize
\centering
\begin{tabular}{@{}lcccccccrr@{}}
\toprule
\textbf{Work} & \textbf{Violations} & \textbf{Legal Def.} & \textbf{Non-CMP} & \textbf{U.I.} & \textbf{P.I.} & \textbf{Subpages} & \textbf{\# V.P.} & \textbf{\# Sites} & \textbf{\# Crawls} \\
\midrule
Our Work (2025) & $\checkmark$ & $\checkmark$ & $\times$ & $\checkmark$ & $\checkmark$ & $\checkmark$ & 8 & 1,793 & 10 \\
Bollinger (2022) & $\checkmark$ & $\checkmark$ & $\times$ & $\times$ & $\times$ & $\checkmark$ & 1 & 29,398 & 1 \\
Rasaii (2023) & $\times$ & $\times$ & $\times$ & $\times$ & $\times$ & $\checkmark$ & 8 & 513 & 5 \\
Kancherla (2024) & $\checkmark$ & $\times$ & $\times$ & $\checkmark$ & $\times$ & $\times$ & 1 & 161 & 1 \\
Matte (2020) & $\checkmark$ & $\checkmark$ & $\times$ & $\times$ & $\times$ & $\times$ & $5^\dagger$ & 1,426 & 1 \\
Sanchez-Rola (2019) & $\checkmark$ & $\checkmark$ & $\times$ & $\checkmark$ & $\checkmark$ & $\times$ & $3^\dagger$ & $2,000^*$ & 1 \\
Liu (2023) & $\checkmark$ & $\checkmark$ & $\times$ & $\times$ & $\checkmark$ & $\times$ & 2 & $100,000^*$ & 8 \\
Van Eijk (2021) & $\times$ & $\times$ & $\times$ & $\checkmark$ & $\times$ & $\times$ & $18^\dagger$ & 603 & 1 \\
Bouhoula (2024) & $\checkmark$ & $\checkmark$ & $\checkmark$ & $\checkmark$ & $\times$ & $\checkmark$ & 1 & 48,843 & 1 \\
Papadogiannakis (2021) & $\checkmark$ & $\times$ & $\checkmark$ & $\times$ & $\checkmark$ & $\times$ & 1 & $27,953^*$ & 3 \\
\bottomrule
\end{tabular}
\caption{Comparison of related work. 
$\dagger$ indicates measurements from vantage points with overlapping legal jurisdiction and 
* indicates inclusion of sites that do not have cookie banners. Ours is the first to measure violations from various legal jurisdictions. Measurement Methodologies: U.I. -- User interfaces, P.I. -- Personal information, V.P. -- Vantage points.}
\label{tab:related-work}
\vspace*{-0.05in}
\end{table*}

Due to the extra-territoriality clause of privacy laws like GDPR and CCPA, 
companies are required to comply with laws so long as they do business with 
any EU or CA resident. Website developers cannot be expected to handle privacy 
compliance with so many regions outside of their own company's location. 
As such, intermediary 3rd-party consent management platforms (CMPs) have standardized compliance for website developers.
Based on the provisions put forth in the various regions worldwide, 
these CMPs convert 
\textit{interpretations} of privacy laws into 
\textit{implementations} of privacy laws.
These CMPs have standardised frameworks for regional privacy laws, directly impacting website developers' 
cookie banner implementations.
CMPs manage users' consent, which are important for compliance 
with GDPR in Europe, PIPEDA in Canada, CCPA in California, and more. 
Hence, CMPs are often implemented such that the geolocation 
from a user's browsing session informs the CMP to display 
a particular cookie banner with its UI and behavior 
customized to the region's regulations~\cite{onetrust2025regulations}. 
For example, in Europe, users more frequently encounter detailed 
consent banners with guidelines set by the GDPR, ePD, and TCF, 
whereas in the U.S., banners may only notify users of data 
collection practices, or simply not appear at all.




\subsection{Related Work}\label{sec:related_work}

Prior studies have focused on three main topics for cookie 
consent: (1) automating or exercising cookie banner UIs, 
(2) analyzing cookie declarations, behavior, and enforcement, 
and (3) legal and usability analyses of consent mechanisms. 

\paragraph{\bf Cookie Banner UI Automation.}
Browser extensions such as Consent-O-Matic and Autoconsent 
automate rejection by matching CSS/JS patterns.
These programs use handcrafted CSS matchers and JavaScript 
object models to create a JavaScript model of consent notices 
and automatically set the cookie banners to rejection.
Ad blockers’ ``annoyance blocking'' closes/hides banners
but does not actively reject cookies. 
Sweepatic’s platform detects non-consensual cookies but only 
before a consent choice being made~\cite{sweepatic_sweepatic_2022}. 
Several researchers study the effectiveness of these automated opt-outs 
such as Khandelwal \etal~\cite{khandelwal_cookieenforcer_2022} which
model consent dialogs as seq2seq tasks and
Demir \etal~\cite{demir2024large} which compares 6 tools 
over 98k pages from 30k sites. 

\paragraph{\bf Cookie Declarations, Behavior \& Enforcement.}
Cookie compliance has been studied since the GDPR's enforcement in 2018. 
Sanchez-Rola \etal~\cite{sanchez-rola_can_2019} manually analyzed 
2,000 sites across EU, USA, China, and other regions, reporting 
that 92\% perform tracking before consent and only 2.5\% erase 
cookies after their placement. They also report that cookie‐settings 
interfaces appear on just 16\% of EU sites and 12\% of US sites. 
Matte \etal~\cite{matte_cookie_2020} also analyzed 1,426 EU sites 
using IAB TCF banners, finding that 54\% contained violations 
like pre‐selected options, consent stored pre‐choice, etc. 
These findings were extended to 3rd-party cookies as well by
Kancherla \etal~\cite{kancherla2024johnny}, who inspected 1,200 
sites (top 200 for UI, 1,000 random from top 5k), reporting that
74\% do not inform third parties of rejection. 
Cookies that remain on the browser were also found to be continually 
used even after rejection. 
Liu \etal~\cite{liu_opted_2022} compared advertiser bidding
before and after GDPR/CCPA opt‐out (across 4 CMPs and 34 IPs) 
and found negligible differences in tracking.

\paragraph{\bf Scaling Cookie Violation Measurements.}
Bollinger \etal~\cite{bollinger2022automating} and Bouhoula 
\etal~\cite{bouhoula2023automated} focused on CMP compliance, defining 
8 violation types (e.g., undeclared cookies and implicit consent) and 
scaled up compliance testing to 30k and 97k sites in the EU, respectively. 
The former found at least one violation in 94.7\% of sites,
with 82.5\% sites containing undeclared cookies and 69.7\% sites 
with implicit consent, and the latter similarly found 65.4\% 
were likely to collect data despite negative consent.

\paragraph{\bf Cookie Banners Across Regions.}
Measurement studies have also been conducted on cookie banners 
in regions world-wide.
Van Eijk \etal~\cite{eijk_impact_2019} crawled 1,543 sites from 
18 countries (27,488 measurements in total) and found that the odds 
of seeing a cookie banner are 102\% higher in the EU. 
Rasaii \etal~\cite{rasaii2023exploring} examined 518 sites across 8 
locations (Sweden, Germany, US West/East, Brazil, South Africa, 
Australia, India), finding cookie banners in 47\% in EU vs.\ 30\% non‐EU.

\paragraph{\bf Legal Requirements \& Usability Studies.}
Santos \etal~\cite{santos2019cookie} identified 22 legal 
requirements for valid consent (e.g., free, specific, informed, prior, 
revocable) and showed that no CMP satisfies them all. 
Santos \etal~\cite{santos2021consent} also determined that CMPs act 
as both data controllers/processors, noting that HTTPS-based CMP s
cripts process IP addresses (an identifier) and shape users’ choices. 
Bouma-Sims \etal~\cite{bouma2023us} performed a usability study of 
OneTrust’s UI across US and UK participants, finding that over half 
could not make a properly informed consent decision 
(52.4\% US vs.\ 46.2\% UK). 

\paragraph{\bf Comparison of \sys with Related Work.}
Unlike prior studies that focus on a single region, law, or CMP, we  
measure cookie consent behavior across 8 jurisdictions, 3 major CMPs, 
and websites from the Tranco top 20k. 
By analyzing legal regulations and CMP documentation, we identified the
root causes that result in cookie banner/consent/placement disparities 
and violations. We conducted measurements at the individual cookie level, 
identifying cookies that likely contained personal information. 
\sys is the first to compare cookie consent violations across 
multiple non-EU regions with privacy laws. 
\Cref{tab:related-work} provides a summary.

\section{Definitions and Terminology}\label{sec:methods}

We provide formal definitions of cookies, consent preferences, and cookie consent violations in \Cref{tab:definitions}.

\begin{table*}[t]
    \footnotesize
    \centering
    \begin{tabular}{l|l|l}
    \toprule
    Term & Definition & Logical Description \\
    \midrule
    Cookie & Defined by (name, $n$; domain, $d$; path, $p$) & $c = \{n,d,p\}$ \\
    Cookie Equivalence & Two cookies (name, $n$; domain, $d$; path, $p$) match & $c_1 \equiv c_2, \iff n_1 = n_2 \land d_1 = d_2 \land p_1 = p_2$ \\
    Cookie Consent Preference & Pair of cookie and consent choice & $P = \{c_i, s_i\}$ where $s_i \in \{$\textit{consent, $\neg$consent}$\}$ \\
    Approved Cookies & Accepted & $A_c = \{ c | (c, s) \in P \land s = consent \}$ \\
    Rejected Cookies & Rejected & $R_c = \{ c | (c, s) \in P \land s = not\_consent \}$ \\
    Compliant Cookie Use & Cookie used consistently with consent choice & $Compliant \iff c \in A_c \land c \notin R_c$ \\
    Ignored Cookie Rejection Violation & Cookie used inconsistently with consent choice & $IgnoredReject \iff c \notin A_c \land c \in R_c$ \\
    Undeclared Cookies Violation & Cookie used without appearing in cookie library & $Undeclared \iff c \notin A_c \land c \notin R_c$ \\
    Ambiguous Cookie Category Violation & Cookie in a rejected + accepted category & $Ambiguous \iff c \in A_c \land c \in R_c$.\\
    \bottomrule
    \end{tabular}
    \caption{All definitions relevant to cookie consent and cookie consent violations.} \label{tab:definitions}
    \vspace{-.4cm}
\end{table*}




\paragraph{\bf Cookie Banners and Cookie Preference Menus.}

``Cookie banners'' inform users of the use of cookies (providing 
no option or confirmation only) and allow users to accept, 
reject, or manage their cookie consent preferences~\cite{degeling_we_2019}. 
Cookie banners can appear as pop-ups on landing pages or can be 
activated by buttons placed in footers or cookie/privacy policy pages. 
These banners are often hidden and must be manually activated
in regions where privacy laws do not require
consent before data collection.
Oftentimes, these banners allow users to open separate ``cookie 
preference menus'' with controls that allow users to set 
their consent preference on specific cookie categories or individual 
cookies (Figure.~\ref{fig:cookie_setting_example}).
The categories can be specified by purposes 
and/or by advertising vendors. 

Websites collect users' cookie consent preferences and 
block/unblock cookies locally based on the consent choice. 
CMPs create and maintain 
special cookies to record users' 
consent choices on websites. For example, \onetrust 
and \cookiebot store users' consent preference in 
cookies named \textit{OptanonConsent}
\cite{onetrust_llc_onetrust_2021} and 
\textit{CookieConsent}~\cite{cybot_duration_2020}, 
respectively. Other sites use 1st-party cookie banners instead of CMPs.

\paragraph{\bf Cookie Definitions.}
Considering a cookie in a browser's cookie storage as a 
tuple of key--values, \sys distinguishes cookies by their 
\textit{name}, \textit{domain}, and \textit{path}. 
This distinction of cookies follows the storage model in 
the cookie specification~\cite{barth_http_2011}. Other 
cookie attributes, while still important to monitor, 
may change over time even for the same cookie (e.g., 
\textit{value} and \textit{expiration time}).

A {\em cookie} represents a value stored on users' browser that 
can be transferred to a website's server. This data collection of
cookies is typically {\em consistent} with user's indicated consent 
preference if the user approves and does not reject the cookie. 
A website correctly enforces a user's consent preference if all 
of the cookies used on the website are consistent with the users' 
cookie consent preferences. A \textit{cookie consent preference} 
is the preference indicated via a user's consent decision to 
a cookie or category of cookies' usage 
(e.g., approve or reject advertising cookies).

\paragraph{\bf Cookie Consent Violations.}
We selected three main forms of cookie consent violations
to investigate in this paper. These violations are defined as 
when cookie consent mechanisms are not consistent in their 
expected/specified behavior and may mislead users to believing 
that tracking cookies are not placed on their browsers. 
These three violations are originally derived by us but are similar 
in nature and scope to the cookie consent violations explored 
in Bollinger \etal's work~\cite{bollinger2022automating}.

An \textit{Ignored Cookie Rejection} violation occurs when 
the website uses a cookie that is rejected and not 
consented by the user. This behavior is misleading, making users 
unaware that cookies are still being used to collect their data 
even after they rejected cookies. Using explicitly rejected 
cookies could violate a user's consent preference.

\textit{Undeclared Cookies} prevent users from setting their 
consent preference for a cookie, so the use of such cookies 
circumvent the principle of consent being freely-given and a choice. 
Undeclared cookie consent violations occur when cookies 
that are not disclosed by the CMP or cookie banner appear on 
the user's browser. The Undeclared Cookies violation is 
different from the \textit{unclassified} cookie category 
(frequently used by CMPs like \cookiebot) which still informs 
users of the usage of the cookies in this category. 

A \textit{Wrong Cookie Category} violation is when the cookie's 
consent is accepted and rejected at the same time. 
This occurs when the categories of cookie banners overlap, 
allowing a cookie to be both rejected and approved by a user. 
For example, a cookie could be included in both the Strictly 
Necessary and Targeting categories. 
A user could have rejected targeting cookies but accepted 
strictly necessary cookies. This is akin to contradictory 
statements in privacy policies identified by the FTC and 
prior studies~\cite{PolicyLint_2019,purpliance_bui_2021,
pitofsky_privacy_2000}.

The existence of Ignored Cookie Rejections and 
Undeclared Cookies seemingly violate the GDPR, as 
Recital 32 states that the conditions for consent require that 
``consent should not be regarded as freely given if the data 
subject has no genuine or free choice or is unable to refuse 
or withdraw consent without detriment''. In Canada, PIPEDA 
guidelines state that consent is reasonable for online behavioral 
advertising if individuals are aware, informed of these practices 
and are easily able to opt out. 
The guidelines also state ``If an individual is not 
able to decline the tracking and targeting...then 
organizations should not be employing that type of technology for 
online behavioral advertising \cite{pipeda2021tracking}.'' 
Similar articles exist in other privacy frameworks like 
PDPA and others.


Note that while these cookie consent mechanisms may be inconsistent with
privacy laws and guidelines, they may not violate the laws. 
Rather, they demonstrate inconsistent behavior from the implementations 
provided by CMPs. In reality, ``consent'' is broadly defined and 
may differ from region to region (e.g., GDPR allows 5 other lawful 
bases beyond consent). 
A thorough review of prior legal cases and court rulings 
may be required to determine whether these ``violations'' 
actually breach any laws or regulations.

\section{Crawler System Design}\label{sec:design}

\subsection{Overview}

Detecting cookie consent violations requires automatically setting 
cookie consent and activating the cookie banners and consent 
preference menus. This remains challenging due to the diversity 
of HTML implementations of consent banner menus. 
Additionally, depending on the region, cookie banners and menus 
may need to be manually activated them. 

Several tools have automated the interactions with 
consent notices~\cite{ghostery_ghosteryautoconsent_2022,
aarhus_university_cavi-auconsent-o-matic_2021,
ninja_cookie_ninja_2022}. 
Khandelwal \etal \cite{khandelwal_cookieenforcer_2022} click 
every element and then detect cookie banner menus after each click. 
Bollinger \etal~\cite{bollinger2022automating} accept/reject 
cookie consent by using the GDPR-specific Consent-O-Matic 
tool~\cite{aarhus_university_cavi-auconsent-o-matic_2021}. 
However, these approaches are less thorough and
scalable than what is needed for our evaluation. 
We wish to audit cookie behavior after setting consent for individual categories, as well as auditing cookie banners that do not pop up.

To remedy this deficiency, \sys activates cookie banner menus by 
using a {\em cookie-button extractor} (\Cref{sec:pref_btn_extract}) 
and a {\em menu activator} (\Cref{sec:activate_cookie_menu}). 
When accessing a web page, if a cookie banner menu is not detected, 
the extractor analyzes HTML elements to extract the candidate 
cookie banner buttons that may activate a menu. 
Cookie consent preferences are extracted in two steps: 
(1) cookie-category consent and (2) cookies in each category. 
Finally, cookie consent violations are detected and classified 
(\Cref{sec:cookie_violation_classifier}).

\subsection{Implementation Details}
\label{subsec:cookie_decl_mapping}

To extract the set of cookies during a website visit, 
\sys extracts the cookies sent to servers via the network 
debugging functionality of Chrome DevTools Protocol that 
reports all HTTP(S) requests with associated 
cookies~\cite{google_inc_chrome_2021}. Dynamic analysis 
is advantageous because it reveals ``real'' occurrences of 
cookies rather than finding only potential ones, thus 
reducing false positives. For example, a website may block 
the use of 3rd-party cookies by preventing the loading 
of 3rd-party scripts and frames without removing the cookies. 
This additional analysis overcomes the limitations of prior 
works that extract all cookies in the 
browser regardless of whether cookies were actually 
transferred to servers~\cite{sanchez-rola_can_2019}.

We posit that a website should enforce a cookie consent 
preference and remove/prevent cookie placement if 
(1) the consent choice is recorded in the browser, 
(2) the cookie matches, and (3) the web page being 
visited is within the scope of consent. 
These conditions ensure the consent preference of a cookie 
applies only when the user browses the web pages to which 
the user gave their consent preference. For example, 
the user's consent preference given to website \textit{a.com} 
does not apply to website \textit{b.com}, although it applies 
to subdomains like \textit{subpage.a.com}. 
\sys maps the cookie declarations in the cookie banners 
to the cookies used by the website by matching cookie 
names and domains. For patterns such as 
\textit{\_gaxxx} or \textit{\_ga\#\#\#}, we assume an `x' 
or `\#' to match any single character. 
However, a single `\#' at the end of the declaration matches 
any alpha-numeric string. For example, the declared cookie 
name "\_gatxxx" matches cookie "\_gat123".
The scope of a consent preference is defined by the domain 
of the consent cookie (such as \textit{OptanonConsent} of 
\onetrust) or other local storage object which records the 
user's consent.
Similarly, the domain names of a cookie and a declaration 
match each other if the declared domain is a suffix of the 
cookie domain. This domain matching scheme is similar to 
the standard cookie domain matching 
specification~\cite{barth_http_2011}.


\begin{figure}[t]
    \centering
    \includegraphics[width=\linewidth,trim=208 156 139 120,clip]{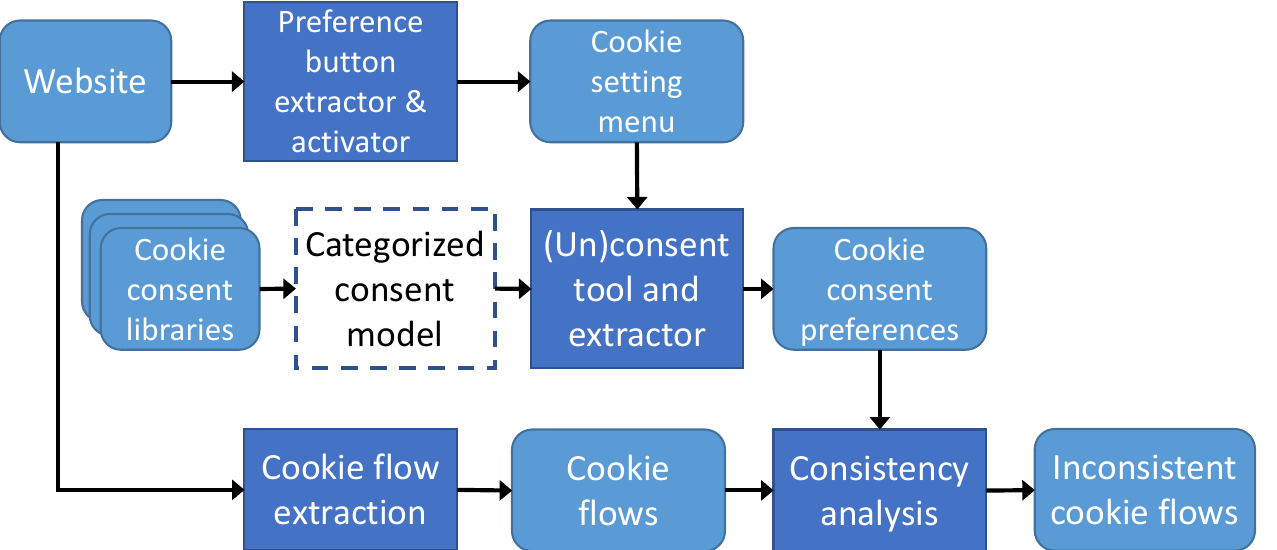}
    \caption{\sys analyzes a website by activating the cookie preference menu and auditing cookie behavior. 
    The dashed box represents an one-time manual step that 
    creates a reusable consent setter for each consent 
    library.}
    \label{fig:system_workflow}
    \vspace*{-0.05in}
\end{figure}

\subsection{Preference Button Extractor}\label{sec:pref_btn_extract}

To increase the coverage of \sys beyond websites 
using consent libraries, we create a detector for 
extracting cookie banner buttons. 
We define a \textit{cookie banner button}
to be an HTML element that, upon 
a user click, displays a menu for the user to set the 
cookie consent. For each web page, the cookie-button 
extractor finds visible HTML elements that represent a 
button or link in all \textit{iframes} contained in the page. 
We consider \textit{a}, \textit{button}, 
\textit{div}, and \textit{span} elements which are 
commonly used to represent links and buttons
\cite{chris_coyier_make_2021,
will_bontrager_software_llc_linking_2021}. 
For \textit{div} elements, we only select leaf elements to 
avoid unrelated elements. \sys uses a random forest model 
with feature groups such as $n$-grams and keywords from HTML 
attributes such as aria-label, class, id, and inner text. 
To increase its coverage, the activator 
attempts to click the top \textit{k} candidate buttons. 
Specifically, if clicking a non-cookie button navigates to 
another page or activates no consent preference menus, 
\sys returns to the initial URL and tries other buttons.
Additional details regarding the datasets, annotation process, 
classification features, performance metrics, and models 
used for our cookie button extractor can be found in 
Appendix~\ref{sec:pref_btn}.

\subsection{Cookie Consent Exerciser}
\label{sec:activate_cookie_menu}

Since CMPs commonly group consent settings into 
cookie categories to simplify the consent process, 
we derive a \textit{categorized consent} analysis 
framework that groups cookies into categories and provides 
the list of cookies of each group. We divide (by purposes 
or vendors, for example) the set of cookies used on a website 
into subsets, called {\em cookie categories} $t_k$. 
A consent preference of cookie category $t_k$ applies to 
all cookies in that category. For example, the consent 
rejection of \textit{krxd.net} domain cookie category
applies to all cookies from that domain. 

\paragraph{\bf Automatic (Un)consent Tool.}
To analyze a specific cookie banner instance, its UI 
controls need to be mapped to the components in the 
analysis framework. The main manual effort is to map the 
HTML elements to the corresponding cookie consent 
categories. We use the Chrome \devtools to 
identify CSS selectors that uniquely identify UI elements 
on the layout. Although the identification of the mapping 
is done manually, we need this manual mapping only once for 
each of the limited number of cookie banners' layouts. 

\paragraph{\bf Cookie Consent Preference Extractor.}
After setting the user's consent preference on the UI, to extract the 
consent preference of each cookie recorded by the cookie library, 
we extract the consent preference for each category and the list of 
cookies for each category. Combining these two lists, 
we get the consent for each individual cookie. 
For example, \onetrust stores consent preferences of categories in the
\textit{OptanonConsent} cookie and the lists of cookies 
per cookie category in \textit{en.json}.
Compared to Consent-O-Matic, which does not support extraction 
of states that requires UI inputs, our tool can measure 
previously unmonitorable inconsistencies of \onetrust. 

\subsection{Cookie Consent Violation Classifier}
\label{sec:cookie_violation_classifier}

Each cookie is then monitored and classified using the logical 
rules set in \Cref{tab:definitions}. Cookies that are placed 
even after the tool rejects cookies are classified as Ignored Cookie 
Rejection Violations. Cookies placed without being declared in the 
cookie library set by the CMP are Undeclared Cookie Violations. 
These violations are classified based on the actions performed by 
the cookie consent exerciser and the cookies declared in the 
CMPs' cookie libraries.

\subsection{System Evaluation}\label{sec:evaluation}

We perform a small-scale end-to-end evaluation of \sys's 
detection of cookie consent violations
and measure the occurrences of each cookie-violation type. 
We also analyze the performance of \sys for ambiguous 
mappings from cookie declarations to browser cookies.

We evaluate the precision of detecting Ignored Cookie 
Rejection, Undeclared Cookies, and Wrong Cookie Category 
violation types by manually rejecting cookie consent on 
the websites. We randomly selected validation sets of 
\lyx{40}, \lyx{40}, and \lyx{30} websites with each of these 
violation types detected in a crawl of the top 20k websites. 
We chose to evaluate a small subset 
of websites with violations since each site required 
manual verification. We visited each website, rejected 
cookies, and visited sub-pages using a clean Chrome 
browser instance. The accepted/rejected cookie 
consent preferences and the transferred cookies were 
recorded by using the \devtools network monitor.


To corroborate the correctness of \sys's detection, we checked 
whether each rejected cookie with Ignored Cookie Rejection 
discovered by \sys was captured in the manual browsing. 
Similarly, we checked whether cookies with Undeclared Cookies 
were unspecified in the cookie banners. 
We manually checked and discussed ambiguous cases (e.g., 
cookie names were declared as \textit{\_ga\_\#}), to 
determine whether the detection was correct.
Using the annotations, our results 
demonstrated that the rule-based detection pipeline has a 
low false positive rate. We manually reproduced and 
verified that \lyx{92.1\%} of cookies with Ignored Cookie 
Rejection, \lyx{91.2\%} with Undeclared Cookies, 
and \lyx{87\%} with Wrong Cookie Category were detected by \sys. 
We checked the correctness of the detected cookie consent violations 
using the CMP and sites' storage objects, such as \textit{en.json}. 
This way, we verified all of the correctly detected violations.

This roughly 90\% FPR from manual validation or reproduction likely 
resulted from some violating cookies not loading
(e.g., embedded YouTube video). 
\sys randomly navigates to subpages for loading cookies. 
We clicked on some random subpages in our manual reproduction, 
but found the subpages and cookies loaded are not deterministic. 





\section{Measurement Methodology}
\label{sec:methodology}

\begin{figure}
    \centering
    \includegraphics[width=0.9\columnwidth]{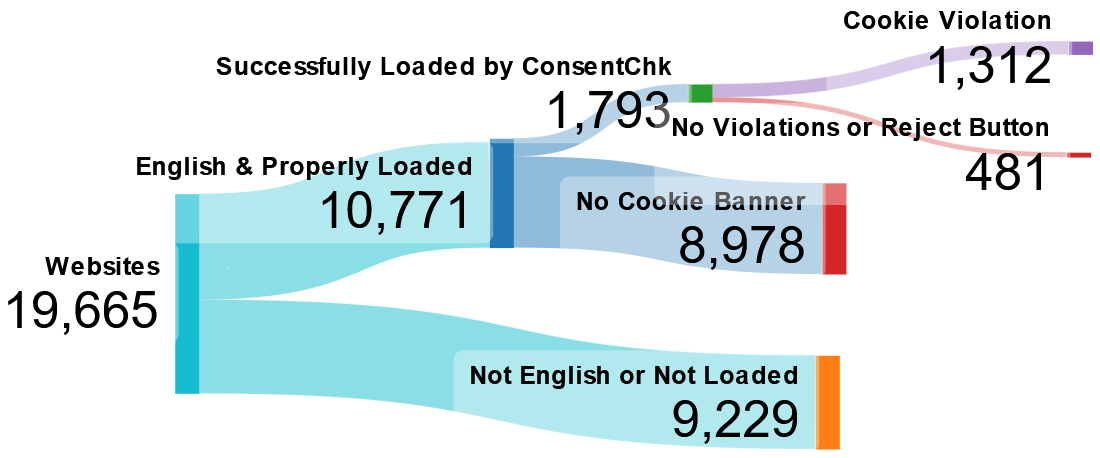}
    \caption{
    Breakdown of sites evaluated in our study.
    }
    \label{fig:sankey_website_flow}
    \vspace*{-0.05in}
\end{figure}

\begin{table}[tbh]
    \footnotesize
    \centering
    \begin{tabular}{lccc}
    \toprule
        CMP  & Market Share & Cookie List? & Cookie Decl.? \\
    \midrule
        Osano            & 2.18\%  & \cmark & \xmark \\
        \textbf{OneTrust}         & \textbf{1.98\%}  & \cmark & \cmark \\
        CookieYes for WP & 1.30\%  & \xmark & \xmark \\
        WP CookieNotice     & 1.21\%  & \xmark & \xmark \\
        \textbf{Cookiebot}        & \textbf{1.20\%}  & \cmark & \cmark \\
        IAB Europe TCF   & 1.15\%  & \xmark & \xmark \\
    \bottomrule
    \end{tabular}
    \caption{The most popular CMPs with more than 1\% market share
        on the top 1M websites 
        as reported by BuiltWith~\cite{builtwith_privacy_2022}.
        }
    \label{tab:builtwith_cmps}
    \vspace*{-0.05in}
\end{table}

\vspace*{-0.05in}
\paragraph{\bf Selected CMPs.}
We report results from the five categories defined by 
cookie libraries we selected --- \onetrust, CookiePro, 
and \cookiebot --- which are some of the most 
widely used on the Web.
Each of \onetrust and \cookiebot has a $>$1\% market share 
while CookiePro has 0.23\%. These CMPs were chosen due to 
their inclusion of both cookie lists and cookie declarations.
Table~\ref{tab:builtwith_cmps} shows the market shares 
and the satisfied criteria of the CMPs.
Since CookiePro is part of \onetrust, we combine it 
into the reported results of \onetrust.

The selected CMPs support different cookie categories. 
The four commonly-supported categories are 
\textit{Necessary}, \textit{Functional}, 
\textit{Analytics}, and \textit{Targeting}.
While \cookiebot use four fixed cookie categories, 
\onetrust supports varying cookie categories.
For example, \textit{scientificamerican.com}
uses \textit{Social Media Cookies}, a customized 
category of \onetrust.
See Appendix~\ref{appd:decode_consent_resources}
for the decoding of consent choice cookies.

\paragraph{\bf Website Selection.}
From the top \lyx{\evalNumSitesCookieSettingStudy} global 
websites in the Tranco list \lyx{November 2023 (ID: 5Y3LN)},
we select \lyx{\nt{\nScannedSites}} websites,
which have an English homepage and were loaded 
successfully, for further analysis with \sys.
Some websites in the list failed to load for various reasons. 
For example, some URLs are non-website ad-serving domains.
The language of the websites is determined by 
a neural-network-based language detector after
converting the web pages to plain text
\cite{google_inc_compact_2021,savand_html2text_2021}.

\paragraph{\bf Measurement Locations.}
We evaluate the detection performance in 
regions with privacy regulations that generally 
require user consent before data collection.
We select Ireland, the UK, California,
Michigan, Canada, South Africa, Singapore, and Australia,
as eight measurement locations. These locations were 
selected because (1) the websites are displayed in English and 
(2) they support a privacy framework requiring notices prior 
to data collection (except Michigan, a US 
state without CCPA-like privacy laws). 
We measured the websites from IP addresses by using 
proxies running on AWS and DigitalOcean, 
two major cloud providers.

\paragraph{\bf Repeated Measurements.}
We first performed crawls in all 8 regions on the top 20k websites. 
Close to 1.8k websites across all regions contained a detected 
cookie banner. After this initial crawl, we recrawled 
the union of the sites across all regions 10 times 
(successfully loaded with cookie banners).
Each measurement iteration involved crawling all 8 regions 
and took roughly 8--12 hours to complete, spanning 1 week 
to complete the 10 measurements.

\subsection{Measurement Procedure}

\sys first opens a clean web browser instance and visits 
the homepage of the website under test. It detects a 
cookie banner button to open the cookie banner menu, 
using the preference button extractor.

After detecting the cookie banner button, \sys rejects 
the cookie consent, reloads the URL where the cookie 
consent choices were submitted, and checks the Consent-Enforcement 
Conditions to ensure that the user's consent choice was recorded. 
These conditions are defined by the consent library used 
by the website and specify when a cookie can be set or 
accessed based on the user's consent preferences.

Next, \sys visits other sub-pages that have hyperlinks on 
the homepage with URLs matching the domain of the 
consent cookie to generate cookie traffic and check for 
cookie consent violations. 
These sub-pages are chosen randomly to ensure a representative 
sample of the website's content is analyzed. 
This step is important for capturing additional cookies used 
on subsites \cite{hils_measuring_2020}. 
\sys rejects all rejectable cookie categories. 

\sys uses \textit{k=5} to try the top-5 
cookie banner button candidates as a trade-off between 
coverage and experimentation duration. Each page contains 
an average of \lyx{232} buttons and links, so using only 
the top \lyx{5} links/buttons reduces the experimentation 
time significantly while still achieving a high recall rate. 
Raising \textit{k} forces the system to check all 
\textit{k} buttons on the websites that do not contain any 
cookie banner buttons, which increases the experimentation 
time significantly.

The crawler uses a \lyx{\timeoutSec}-second timeout to 
load the pages. We found this timeout sufficient to 
completely load most of the web pages with the fast 
network of our servers and cloud providers. 
The crawler uses the Playwright browser automation tool
\cite{microsoft_microsoftplaywright-python_2020} to 
control the Google Chrome web browser and utilizes 
techniques provided by an automatic browsing plugin 
to avoid getting detected by bot detection 
(i.e., Puppeteer and Playwright's stealth mode) 
\cite{berstend_puppeteer-extra-plugin-stealth_2021}. These added Playwright functionalities mimic human behavior, bot flagging via headless detection and user agents.

We conducted experiments in a distributed framework 
based on Docker Swarm \cite{docker_inc_swarm_2021} on 
4 machines with 1.08TB RAM and \lyx{96} task queue workers. 
The cookie-consent scanning of the \lyx{\nt{\nScannedSites}} 
websites took \lyx{40} hours to perform the measurements 
from the \lyx{8} locations. The crawls were performed
during October 4--12, 2024.


Of the top 20k websites, 10.8k websites were in English and 
properly loaded. From this subset, \sys found 1,793 websites 
with cookie banners and consent settings from our list of CMPs. 
\sys analyzed the cookie consent behavior of 1,312 sites containing cookie 
preference menus (see the breakdown in \Cref{fig:sankey_website_flow}).
\sys collected information on cookie banner parameters, cookies declared
in the CMP, cookies placed on the browser, website traffic 
and navigated subpages.


\subsection{Personal Information Analysis}
\label{sec:personal_information}

Most privacy laws only pertain to personal information -- that is data 
that could be used to either identify an individual or data pertaining 
to an identifiable individual. As we are primarily interested in studying 
cookie consent violations, we need a method for differentiating 
cookies containing personal information or identifiers from those not.
For example, some cookies may simply contain a number to 
indicate a user's preference for light/dark mode on a site 
while other cookies may contain the latitude and 
longitude of a user or a unique tracking identifier.
The cookie declaration may also include its purposes
that describe the nature of data usage. For example, 
the cookie \textit{loc} with domain \textit{addthis.com} 
collects the location of users to help 
\textit{addthis.com} track their location when its 
share buttons are clicked.
Because of this and the fact that a vast majority of the cookie 
consent violation types are Undeclared Cookies, we need a system 
to detect and categorize whether cookies contain personal information, using the cookie's name, value, and declared purpose.
We built a lightweight detector that searches for keywords 
(e.g., city, postal code, state, country name, and uid) 
and regex patterns (e.g., IP address, GPS location, and 
common tracker formats), as well as decoding Base64-encoded cookies. 
The detector also leverages the zxcvbn~\cite{wheeler2016zxcvbn} library for measuring
the entropy of the cookie values (to discover tracking UIDs).
For those cookie purposes not declared by website developers 
in the cookie libraries, our detector referenced 
Cookiepedia~\cite{one_trust_cookiepedia_2020} as well as 
other open-source cookie databases. 
We opted to use this lightweight approach for detecting 
personal information to scale our approach to millions of cookies.

Roughly 75\% of all cookies found in our measurement likely
involve personal information. 
Regional discrepancies exist as well, with the EU and UK having 
roughly 3\% more cookies including those with personal information
than other regions.
\textit{Unless stated otherwise, the reported results only 
include cookies that were detected to have personal information.}

\subsection{Analysis Methodology}

We analyzed 3 different aspects of cookie banners on the websites 
across the 8 regions: (1) the cookies placed on each website across 
each region; (2) the differences in cookie consent violations 
we detected for each cookie on the websites across each region; 
and (3) the differences in UI settings and parameters for 
the websites across each region. We report both high-level 
trends in cookie consent violations and cookie count as well as 
per-site and pairwise region differences. 
In reporting total violation counts (\Cref{tab:inconsistency}), 
we report the union across the 10 measurements. In other plots 
(\Cref{fig:normalized_cookie_violations,fig:3rd_party_violations}), 
we report the average. \textit{Unless stated otherwise, the reported results use 
Ireland as a baseline due to the strict requirements on cookies from the GDPR.}

\subsection{Statistical Significance Tests}
\label{sec:statistical_significance}

\begin{table}[t]
\centering
\footnotesize
\centering
\begin{tabular}{lrlrl}
\toprule
\multirow{2}{*}{Measurement} & \multicolumn{2}{c}{1st-Party Cookies} & 
    \multicolumn{2}{c}{3rd-Party Cookies} \\
\cmidrule(lr){2-3} \cmidrule(lr){4-5}
             & Levene's & H-Test (p) & Levene's & H-Test (p) \\
\midrule
Mean Cookies & 4e-6 & 69.9 (2e-12) & 0.0023 & 71.7 (7e-13) \\
Ignored Reject & 0.0074 & 73.9 (2e-13) & 0.0348 & 73.3 (3e-13) \\
Undeclared & 6e-5 & 68.0 (4e-12) & 0.0017 & 70.5 (1e-12) \\
Wrong Categ. & 0.0005 & 69.7 (2e-12) & 0.0002 & 57.7 (4e-10) \\
\midrule\\
\end{tabular}
\vspace{-.4cm}
\caption{Statistical Significance Tests for \Cref{fig:cookie_count,fig:normalized_cookie_violations,fig:3rd_party_violations}.}
\label{tab:statistical_tests}
\vspace*{-0.05in}
\end{table}

Our analysis includes statistical significant tests, Levene's 
Test and Kruskal Wallis H Test for testing the (non)homogeneity of 
variances and for comparing multiple independent samples.
In our study, Levene’s Test was applied to both first- and 
third-party cookie counts across the eight regions. As shown in 
Table~\ref{tab:statistical_tests}, all p-values from Levene’s 
Test are less than 0.05 for both first- and third-party cookies, 
indicating the significantly different variances in cookie counts 
and violations across regions (non-homogeneous variances).

We then used the Kruskal-Wallis H Test, a non-parametric test 
that compares the medians of two or more independent groups and 
is suitable when the assumptions of normality are not met, 
which was the case for the distribution of cookies and 
cookie consent violations in the regions we studied. 
We used this test to determine whether cookie placement and 
violation measurements were significantly different across regions. 
Our application of the H test revealed significant 
H-statistics for all variables analyzed 
(Table~\ref{tab:statistical_tests}) in both first- and third-party 
cookies, demonstrating statistically significant discrepancies 
in cookie placement and cookie violation rates between regions.

\subsection{Limitations of Our Study}

The nature of our measurement methodology accompanies the inherent 
biases and limitations that may affect the generalizability 
of our study. These limitations arise from our website/CMP 
selections, embedded content not deterministically loading in 
our random crawls, using commercial server IPs as opposed to 
residential IP addresses, and not studying websites without cookie 
banners/settings loading. Our website selection may affect the 
representativeness of our results, given that our measurements cover 
a fraction of websites (i.e., 20K potential websites and 1.3k that 
display cookie banners). Our intent is not to claim that all websites 
and CMPs exhibit these behaviors, but rather to study the subset 
of sites that deploy popular CMPs.

\section{Measurement Results}\label{sec:study}

\begin{figure*}[t]
    \centering
    \begin{subfigure}[t]{0.33\textwidth}
        \centering
        \includegraphics[width=\textwidth]{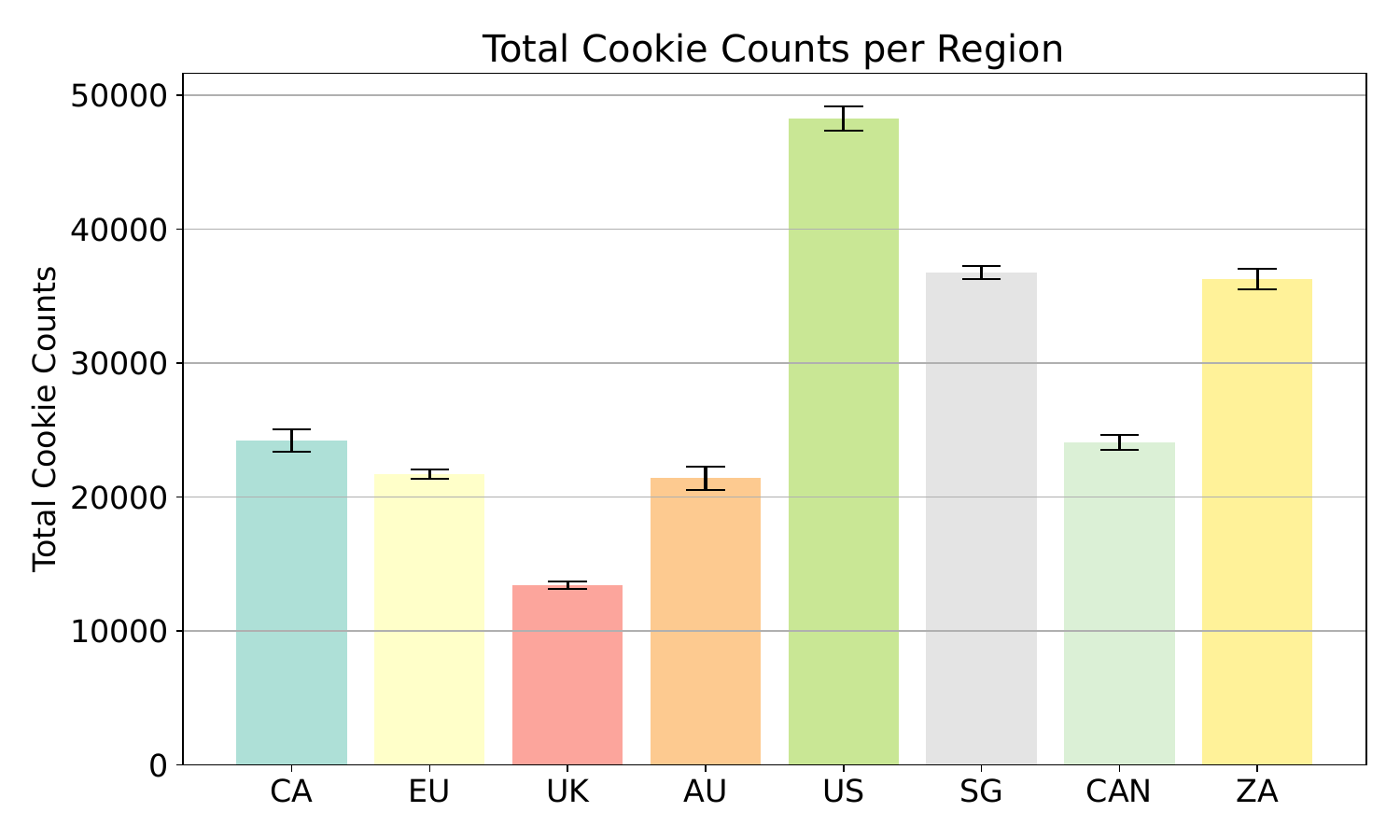}
        \caption{Total cookie counts}
        \label{fig:total_cookies}
    \end{subfigure}
    \begin{subfigure}[t]{0.33\textwidth}
        \centering
        \includegraphics[width=\textwidth]{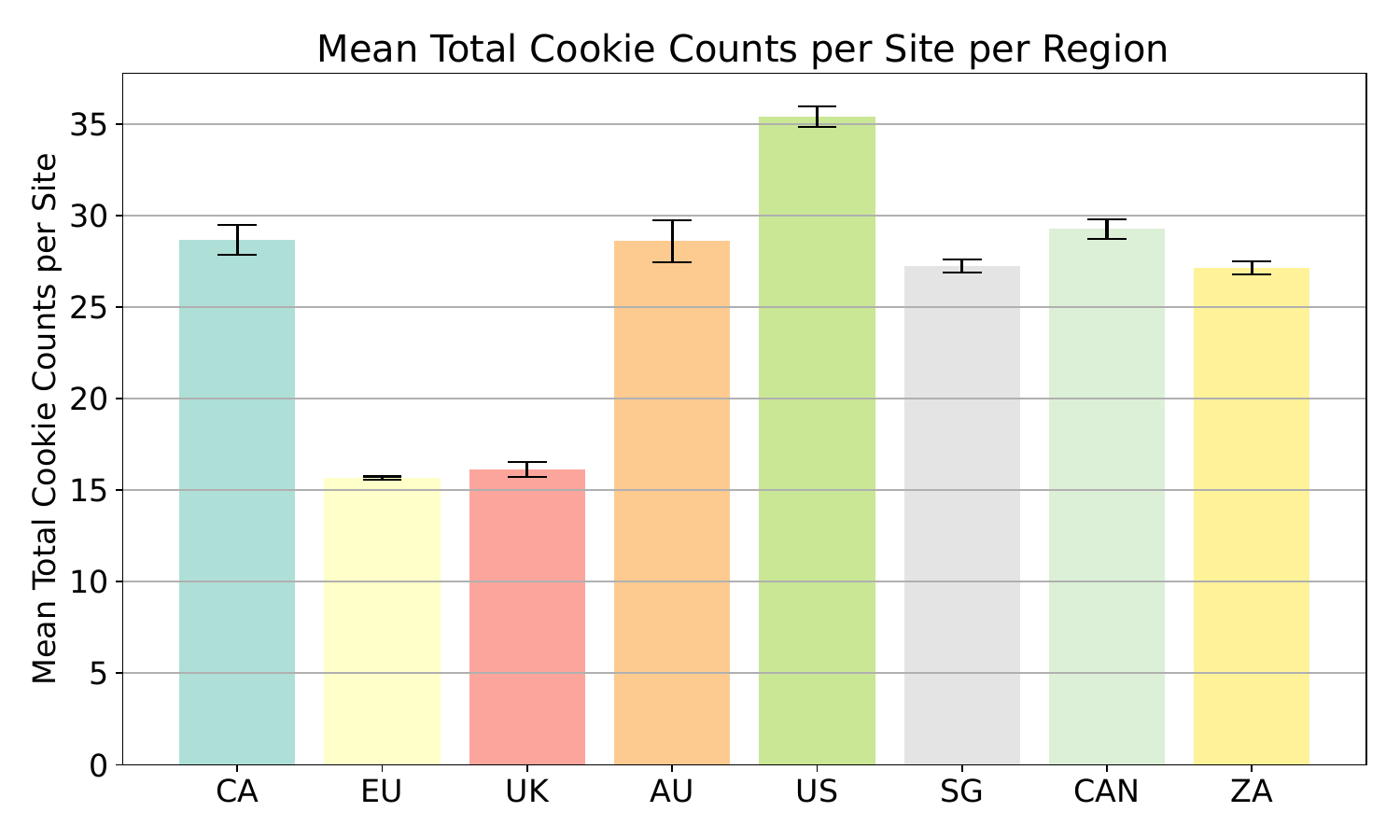}
        \caption{Mean cookie counts per website}
        \label{fig:cookies_per_site}
    \end{subfigure}
    \begin{subfigure}[t]{0.33\textwidth}
        \centering
        \includegraphics[width=\textwidth]{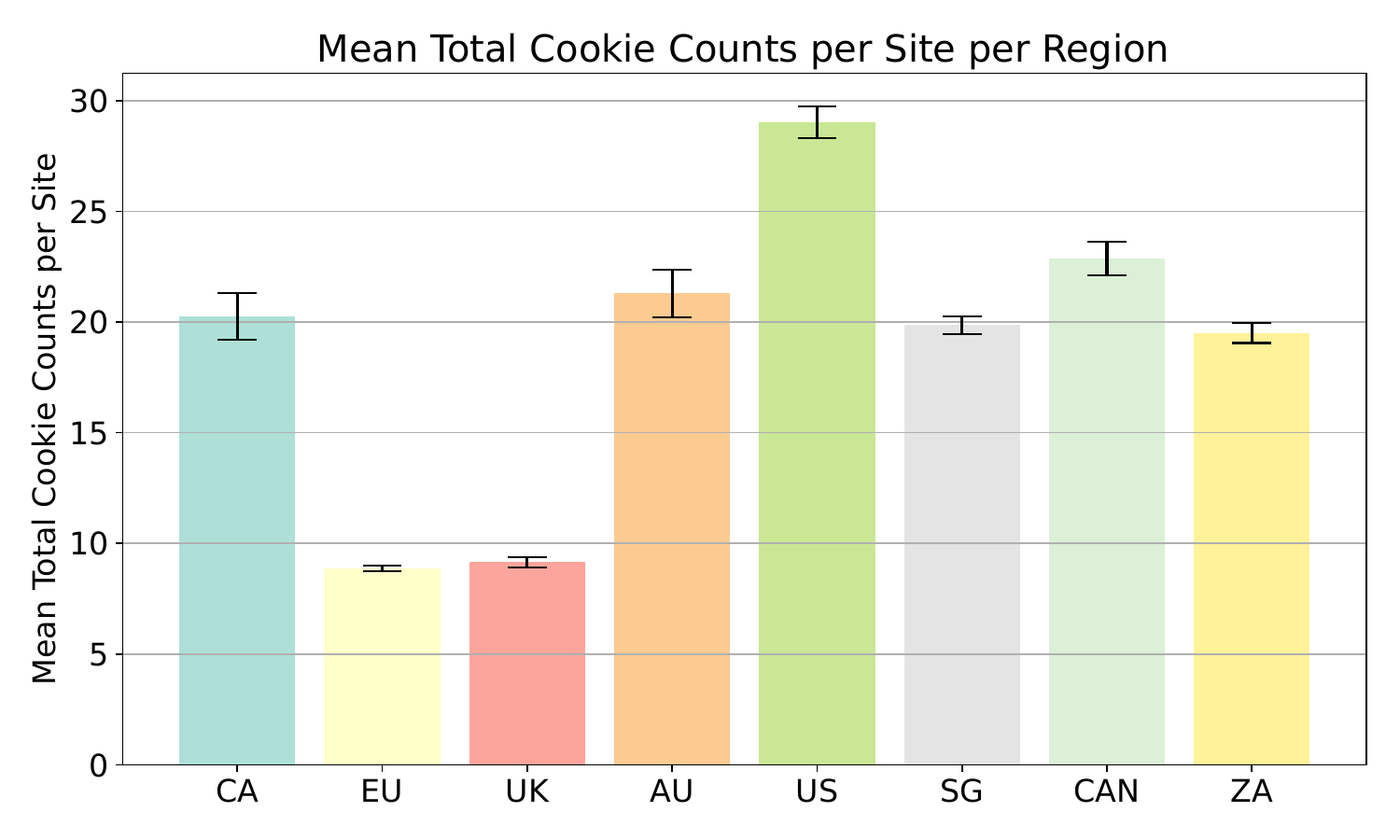}
        \caption{Mean 3rd-party cookie counts per website}
        \label{fig:3rd_party_cookies_per_site}
    \end{subfigure}
    \vspace*{-0.07in}
    \caption{Cookie counts measured in our study.}
    \label{fig:cookie_count}
\end{figure*}




Many websites did not honor users' preferences 
on cookie consent. 
Specifically, we found that 96.18\% of websites accessed 
in the EU contain at least one cookie consent violation. 
We found that undeclared cookies constituted \lyx{47.35\%} of personal information cookies on websites.

We also discovered that most websites failed to remove cookies on consent rejections. 43.12\% of cookies on websites ignored users' cookie consent rejection.

We also detected contradictory cookie banners on 3.13\% of 
websites (Wrong Cookie Category Violation).
These cookies had contradictory consent preferences 
with the same \textit{Name} and \textit{Host} but included 
in two different categories. For example, \textit{\_gid} 
cookie of \textit{cambridge.org} was listed in both 
always-active Necessary and rejectable Performance categories.

Websites made it challenging for users to opt out of 
unnecessary cookies. Despite the Necessary cookie category 
being designed to indicate cookies required purely for website 
services, we found \lyx{\ShowPercentage{\nsiteExtraUn}
{\nSitesWithSettings}} of websites with un-Necessary cookie 
categories set unnecessary cookies as "always active." 
This would prevent users from opting out of tracking 
practices.

Few websites correctly enforced the user's cookie
consent preferences on all cookies. Specifically, only 3.82\% of websites
correctly enforced the consent 
preferences of users.

\subsection{High-Level Observations}
\label{sec:observations}
\vspace*{-0.05in}
\paragraph{\bf Cookie Consent Violations.}
In each of the eight regions, at least one cookie consent violation 
was found to occur in up to \lyx{96.18--97.72\%} of websites. 
The wrong cookie category violation was detected with the least 
prevalence, with \lyx{218--446} cookies. 
The undeclared cookies violation was the most prevalent, with 
\lyx{54,633--250,447} cookies.
\vspace*{-0.05in}
\paragraph{\bf Consent-Ignoring Cookies.}  
The tracking cookies of Google (\_ga and \_gid) and 
Meta (\_fbp) are the most common consent-violators.
This highlights an important fact that users cannot opt out 
of tracking even when they explicitly reject 
the consent of such types of cookies.
Table~\ref{tab:top_incor_cookies} shows the most common 
cookies with an Ignored Cookie Rejection violation.


\vspace*{-0.05in}
\paragraph{\bf Consent-Ignoring Trackers.}
We measured the number of 3rd-party cookies \& trackers 
and the websites they were on.
The most common cookie domains are found to be
\textit{doubleclick.net} and \textit{youtube.com}
which are tracking domains of the same owner \textit{google.com}. 
Table~\ref{tab:top_incor_receivers} lists the top trackers.

\vspace*{-0.05in}
\paragraph{\bf Regional Discrepancies.}
The results of our crawls demonstrate a stark discrepancy 
in cookie count, cookie consent violations, and cookie banner implementations across different regions. These differences
result in up to 3.44\% more websites with undeclared cookies 
in some regions than the baseline EU and UK regions. 
The US contains 250,447, almost $3\times$ more cookies than 
websites in the EU, and as much as 10\% more undeclared 
cookies per website. A large number of websites also contain 
differences in cookie library implementations, with non-European 
regions containing more cookie banners that do not display ``Reject 
All Cookies'' buttons, have shorter consent lifetimes, and 
have opt-out or implied consent rather than opt-in banners. 
\vspace*{-0.05in}
\paragraph{\bf Personal Information.}
We analyze the contents of each cookie found in a cookie 
consent violation to determine whether they contain 
personal information, potentially violating user consent. 
The 75\% of cookie consent violations 
likely involve cookies containing personal information.
\Cref{tab:personal_information} contains statistics on cookies with 
tracking IDs (64.10\%), location (7.44\%), IP address (5.74\%), 
and language (1\%).

\subsection{Major Findings}
\label{sec:findings}

\begin{table*}[t]
\centering
\footnotesize
\centering
\begin{tabular}{lrlrlrlrl}
\toprule
\multirow{2}{*}{Violation Type} & \multicolumn{2}{c}{California (CA)} & 
    \multicolumn{2}{c}{Ireland (EU)} & \multicolumn{2}{c}{United Kingdom (UK)} & \multicolumn{2}{c}{Australia (AU)} \\
\cmidrule(lr){2-3} \cmidrule(lr){4-5} \cmidrule(lr){6-7} \cmidrule(lr){8-9}
             & \# Cookies & \% Websites & \# Cookies & \% Websites& \# Cookies & \% Websites & \# Cookies & \% Websites  \\
\midrule
Ignored Cookie Rejection & 83,036 & 88.41\% & 72,928 & 86.09\% & 44,605 & 84.75\% & 60,985 & 80.74\% \\
Undeclared Cookies & 115,718 & 93.61\% & 87,251 & 90.19\% & 54,633 & 90.04\% & 99,893 & 92.86\% \\
Wrong Cookie Category & 316 & 3.03\% & 365 & 3.13\% & 206 & 2.82\% & 218 & 3.20\% \\
\midrule
\multirow{2}{*}{Violation Type} & \multicolumn{2}{c}{Michigan (US)} & 
    \multicolumn{2}{c}{Singapore (SG)} & \multicolumn{2}{c}{Canada (CAN)} & \multicolumn{2}{c}{South Africa (ZA)} \\
\cmidrule(lr){2-3} \cmidrule(lr){4-5} \cmidrule(lr){6-7} \cmidrule(lr){8-9}
             & \# Cookies & \% Websites & \# Cookies & \% Websites & \# Cookies & \% Websites & \# Cookies & \% Websites  \\

\midrule
Ignored Cookie Rejection & 128,871 & 84.65\% & 107,233 & 82.22\% & 65,882 & 81.63\% & 107,072 & 81.97\% \\
Undeclared Cookies & 250,447 & 94.15\% & 167,979 & 92.90\% & 117,144 & 91.20\% & 165,647 & 93.56\% \\
Wrong Cookie Category & 300 & 2.50\% & 395 & 3.37\% & 272 & 3.35\% & 446 & 3.69\% \\
\bottomrule
~\\
\end{tabular}
\vspace{-.4cm}
\caption{Total detected cookie violations across 10 repeated measurements.}
\label{tab:inconsistency}
\end{table*}

\noindent
{\em \underline{\bf Finding 1}: The US has the most cookies,
undeclared cookie violations, and ignored cookie rejection violations.}
Even in CA, the state with the strictest privacy laws, 
the prevalence of undeclared cookies per site was found to be 
higher than the EU by 5.49\%. The total undeclared cookie and 
ignored cookie rejection violation counts across 10 measurements in the US (MI)
were 250,447 and 128,871 cookies, respectively. 

\noindent
{\em \underline{\bf Finding 2}: Cookies are much more prevalent 
in the non-GDPR regions.}
European countries have the fewest 1st- and 3rd-party cookies 
placed on websites (EU has roughly 15.66 1st-party cookies per site and 
8.88 3rd-party cookies per site, respectively). The US-MI has the most, with 
roughly 35.4 (1st-party) and 29.03 (3rd-party) cookies per site, followed by all other non-EU
regions with between 27.14--29.26 1st-party cookies per site. 

\noindent
{\em\underline{\bf Finding 3}: 3rd-party cookies still constitute 
a large number of cookie consent violations across regions.}
While the EU and UK have significantly fewer 3rd-party cookies 
(\Cref{fig:3rd_party_cookies_per_site}), all other regions have 
a large number of 3rd-party cookies placed on each site. Third-party cookies in the US made up roughly 29.03 cookies per site while websites in the EU and UK contained the fewest with an average of 8.88 and 9.16 cookies per site, respectively. These violations were found to be statistically significant (\Cref{sec:statistical_significance}).

\begin{figure*}[t]
    \centering
    \begin{subfigure}[t]{0.33\textwidth}
        \centering
        \includegraphics[width=\textwidth]{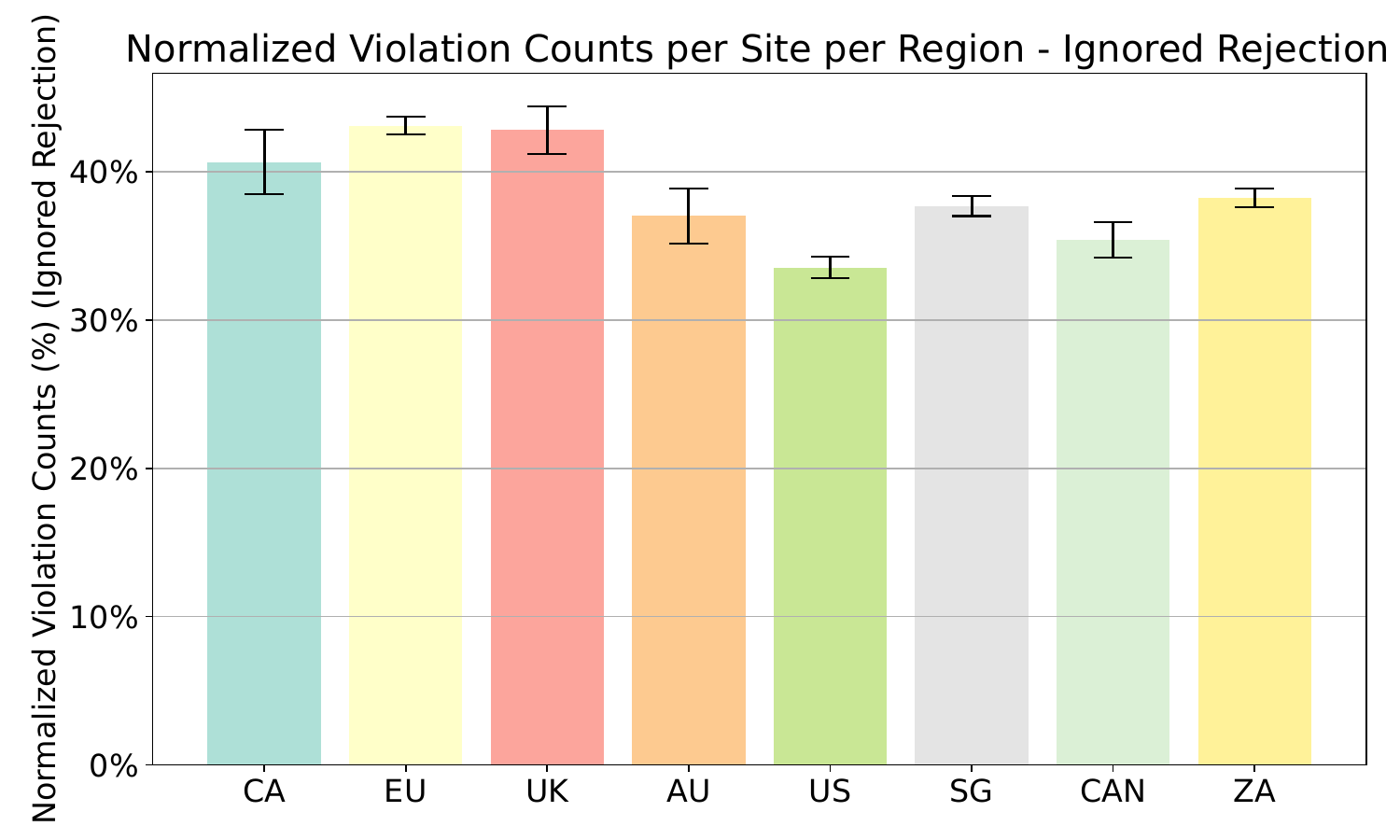}
        \caption{Ignored Cookie Rejection}
        \label{fig:violation_ignored_cookie}
    \end{subfigure}
    \begin{subfigure}[t]{0.33\textwidth}
        \centering
        \includegraphics[width=\textwidth]{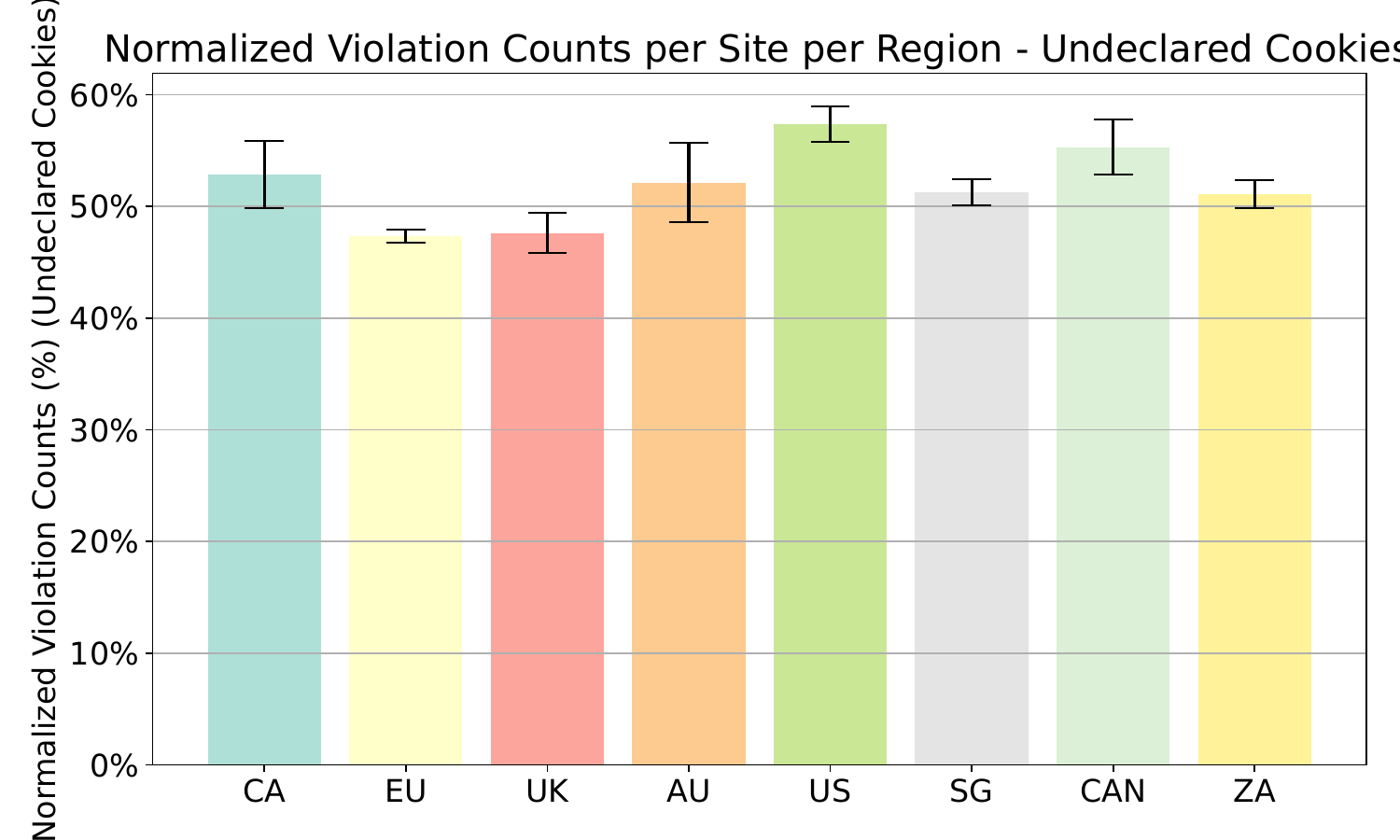}
        \caption{Undeclared Cookies}
        \label{fig:violation_undeclared_cookie}
    \end{subfigure}
    \begin{subfigure}[t]{0.33\textwidth}
        \centering
        \includegraphics[width=\textwidth]{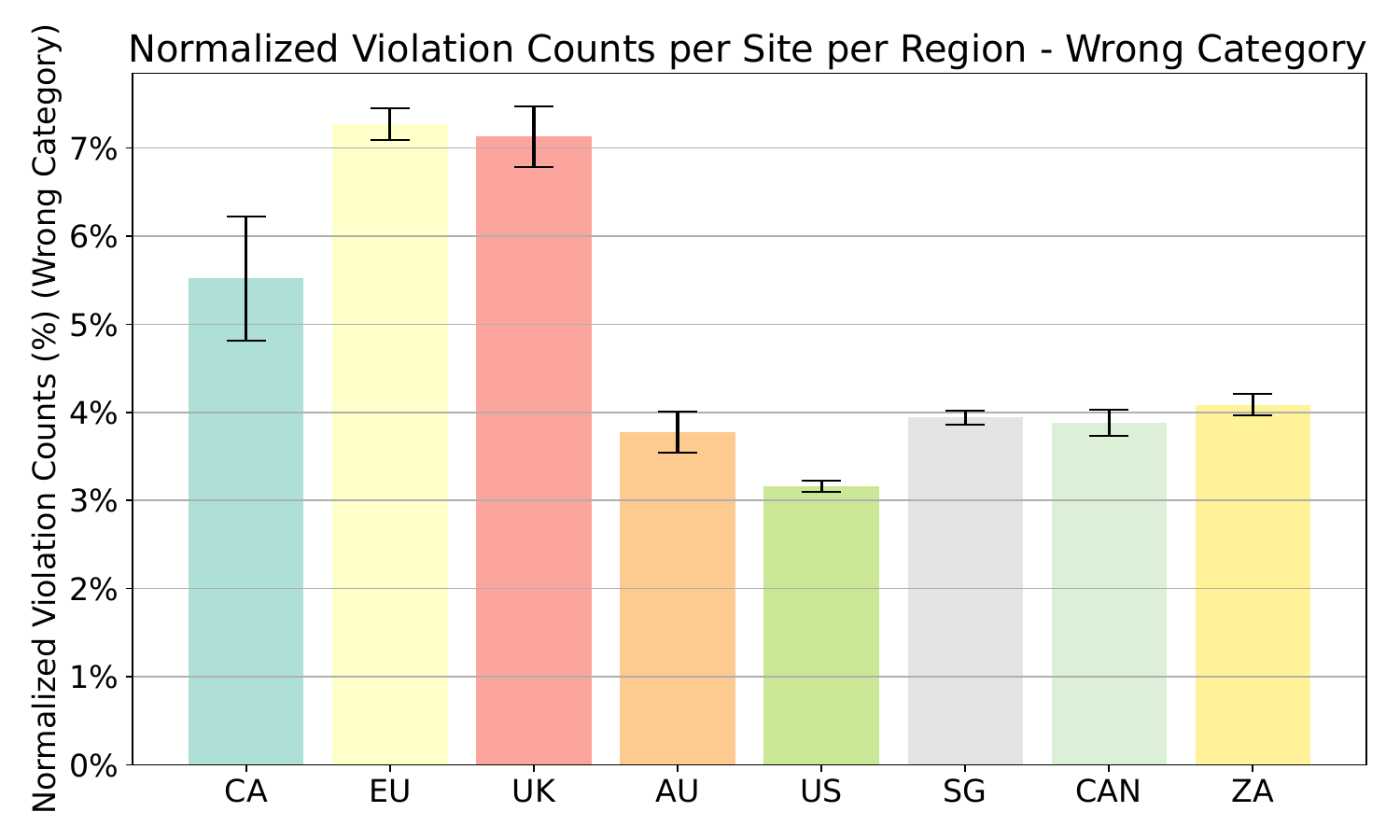}
        \caption{Wrong Cookie Category}
        \label{fig:violation_wrong_category}
    \end{subfigure}
    \vspace*{-0.07in}
    \caption{Normalized average cookie violations per-website.}
    \label{fig:normalized_cookie_violations}
\vspace*{-0.07in}
\end{figure*}

\noindent
{\em\underline{\bf Finding 4}: Regions with fewer undeclared 
cookie consent violations have more ignored cookie rejection and 
wrong cookie category violations.}
\Cref{fig:violation_ignored_cookie} and 
\cref{fig:violation_undeclared_cookie} show an inverse correlation 
between the number of ignored rejection violations and undeclared 
cookie consent violations. Most likely, website developers are more vigilant about categorizing undeclared cookies for GDPR compliance but neglect to update their non-GDPR CMP cookie lists. Consequently, sites in the EU and UK have higher rates of ignored cookie rejections likely due to an increase in declared cookies, but a lack of properly implemented CMP scripts.
Undeclared cookies are placed regardless of rejection. These cookies are not even present in the cookie library and thus cannot be rejected. 

\noindent
{\em\underline{\bf Finding 5}: Cookie placement and violation rates 
can be categorized into three groups}: EU/UK (fewest cookies, 
highest compliance), CA, AU, SG, CAN, ZA (moderate cookie count
and compliance), US (most cookies, lowest compliance). 
Without any privacy regulations, the US has poor cookie practices. 
In our analysis of websites' cookies and cookie consent violations, 
we ran pairwise comparisons on websites between different regions, 
discovering that all other regions, particularly the US, have a 
significant increase in cookie count and cookie consent violations 
compared to the EU and UK 
(\Cref{fig:difference_cookies,fig:difference_cookie_violations}). 
These discrepancies were found to be statistically significant 
(\Cref{sec:statistical_significance}).

\noindent
{\em\underline{\bf Finding 6}: Cookie banners, their UIs, and 
their functionalities are different across regions.}
Across the 8 studied regions, the cookie libraries and cookie 
banners have a significant number of differences in their 
configurations. The EU and US have the most pairwise disparities 
in cookie banner parameters. SG (5084), ZA (5097), and CA (4870) also contain a significant amount of cookie 
banner implementation differences to the EU, while Canada (3,050) and 
Australia (2,705) contain fewer differences. Websites in the EU and UK have more 
privacy-preserving configurations compared to other regions 
(consent models, reject all button appearance, etc.).

\subsection{Contextualizing Findings with Prior Work}
\label{sec:contextualization}

Although every measurement study uses different methodologies, 
we provide a comparison of broad trends to paint a picture of 
cookie consent compliance over time.

\vspace*{-0.05in}
\paragraph{Increase in Overall Violation Rates.} Early audits 
reported very high violation prevalence even on small samples. 
Sanchez-Rola (2019) \etal~\cite{sanchez-rola_can_2019} found 
that 92\% of 2k audited sites set tracking cookies before consent. 
Bollinger (2022) \etal~\cite{bollinger2022automating} observed an increase of violations to 94.7\% of 30k sites. Our crawl found 
at least one violation on 96.18--97.72\% of sites --- an 
absolute increase of 1--3\% every three years. 
Though this increase may be due to many factors: website developers not updating CMPs and repeated measurements catching 
more violation instances, cookie consent violations appear 
to be increasing.

\vspace*{-0.05in}
\paragraph{Undeclared Cookies.}
Bollinger \etal~\cite{bollinger2022automating} reported 82.5\% 
of sites deploying undeclared cookies in the EU. For other non-EU regions, we found that undeclared cookie violations were even more prevalent. Undeclared cookies 
remained the dominant category, with 90.19--94.15\% of sites 
containing at least one undeclared cookie. These undeclared 
cookie violations are found to be more prevalent (3--4\%) 
in non-EU regions than EU regions.

\vspace*{-0.05in}
\paragraph{Implicit Consent and Ignored Rejections.}
Implicit consent —-- often manifested as cookies being set despite 
explicit rejection —-- has persisted and even grown more widespread. 
Bollinger (2022) \etal~\cite{bollinger2022automating} report that 
69.7\% of sites commit implicit‐consent violations and 21.3\% 
ignore users’ reject clicks, while Sanchez‐Rola (2019) 
\etal~\cite{sanchez-rola_can_2019} found 92\% of 2,000 audited 
sites performed tracking before any consent. 
Bouhoula (2023) \etal~\cite{bouhoula2023automated} show 77.5\% of 
EU sites implicitly consent when banners are closed.
We found that “Ignored Cookie Rejections” affect 80.74\% (Australia) 
up to 88.41\% (California) of sites --- with 44,605 ignored‐rejection 
cookies in the UK and 128,871 in the US.

\vspace*{-0.05in}
\paragraph{Geolocation-Specific Disparities.}
Rasaii \etal~\cite{rasaii2023exploring} reported banner prevalence
of 47\% in EU vs.\ 30\% non-EU, and 83-–96\% more tracking cookies 
outside the EU. Eijk \etal~\cite{eijk_impact_2019} measured a
102\% increase in banner odds in the EU. We extend these findings by 
quantifying both violation counts and UI configurations: US sites 
average 46.5 first‐party and 37.3 third‐party cookies per 
site, approximately three times the EU’s 20.0 and 11.0.
Meanwhile, non-EU regions are 3.4\% more likely to omit “Reject All” 
buttons and default to opt-out or implied consent models 
(32\% of US sites).

\subsection{Cookie Consent Violations Across Regions}

Overall, the EU and UK have fewer undeclared cookies (87,251 and 
54,633), websites with undeclared cookies (90.19\% and 90.04\%), 
and undeclared cookie consent violations as a proportion of all 
cookies. Conversely, the number of ignored cookie 
rejection violations when normalized to the total cookie count is 
higher in these regions (\Cref{fig:normalized_cookie_violations}). 
Such a phenomenon can be explained by the increase in proper 
cookie declarations. For a cookie to be classified as an ignored 
cookie rejection by our system, it needs to be 
declared by the cookie library. Thus, the increase in declared cookies 
results in more cookies being continued to be placed even after the user 
rejects all cookies. In contrast, the US had the most websites with 
undeclared cookies with 95.93\% of websites containing at least one 
undeclared cookie violation. AU had the third-fewest cookies loaded 
in our measurements. The region also had the fewest websites with 
ignored cookie rejection violations with only 80.74\% containing 
these violations. CA had the most websites with at least one 
ignored cookie rejection, with 88.41\%. 
Across all regions, wrong cookie category violations were found 
on very few websites, between only 2.50\% and 3.69\% of websites had one 
violation of this type.
Despite Canada's PIPEDA having similar guidelines as the EU/UK 
regarding tracking cookie consent, this region contained more cookie 
consent violations. nhl.com contains cookies from googleadservices.com, an example of such an undeclared cookie violation (for tracking cookies in Canada).
We found from our measurements that the same websites had many more 
cookies and cookie consent violations when accessed in 
different regions. 
Websites in non-EU/UK regions had roughly 8 additional cookies 
(\Cref{fig:difference_cookies}) and 7 additional cookie consent 
violations (\Cref{fig:difference_cookie_violations}). 
As an example, nvidia.com had 16.25--24.00 cookies in AU, SG, EU, UK, CAN whereas it had 28.86--36.78 cookies in CA, ZA, US.

\subsection{3rd-Party Cookies and 
Consent Violations}

\begin{figure*}[t]
    \centering
    \begin{subfigure}[t]{0.33\textwidth}
        \centering
        \includegraphics[width=\textwidth]{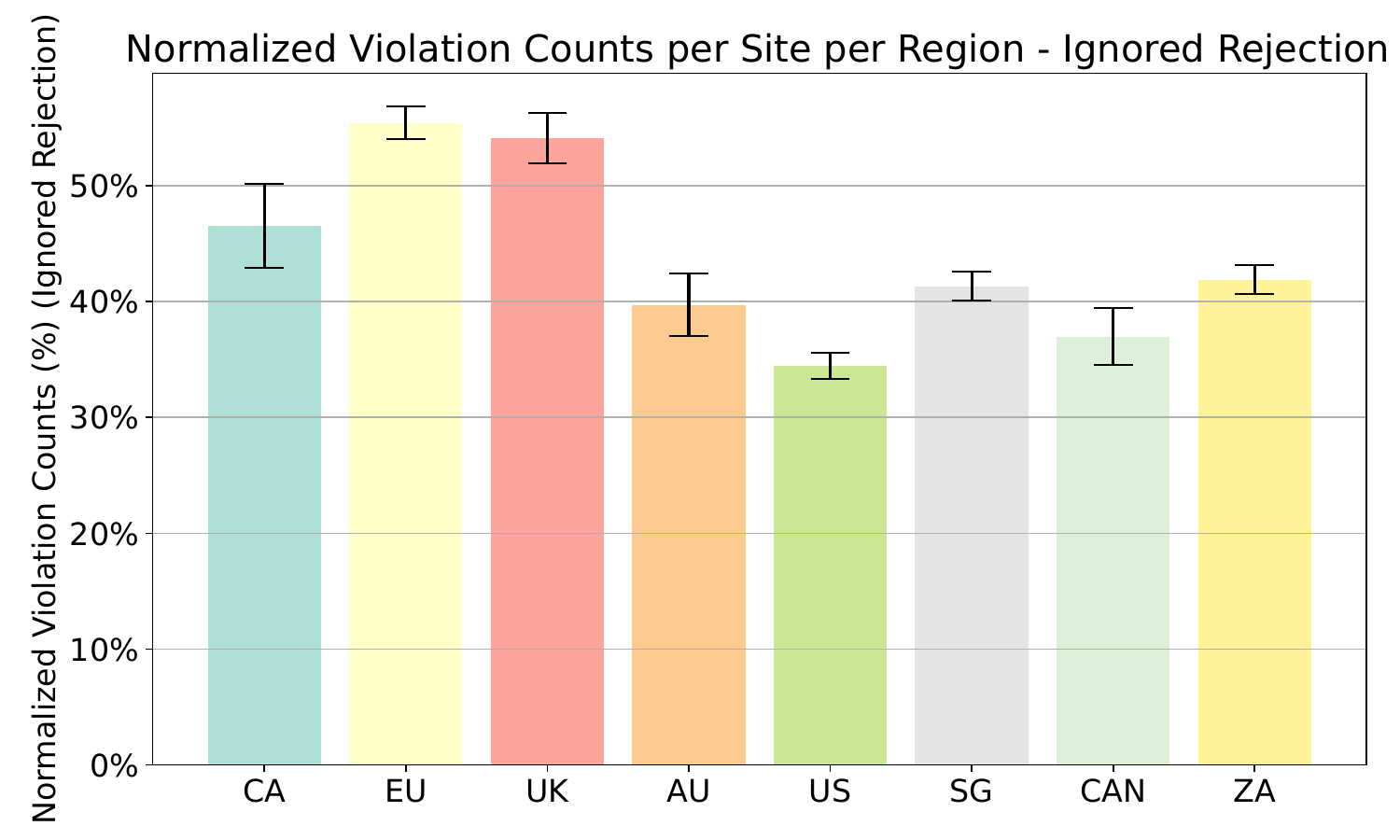}
        \caption{Ignored Cookie Rejection}
        \label{fig:3rd_party_ignored_cookie}
    \end{subfigure}
    \begin{subfigure}[t]{0.33\textwidth}
        \centering
        \includegraphics[width=\textwidth]{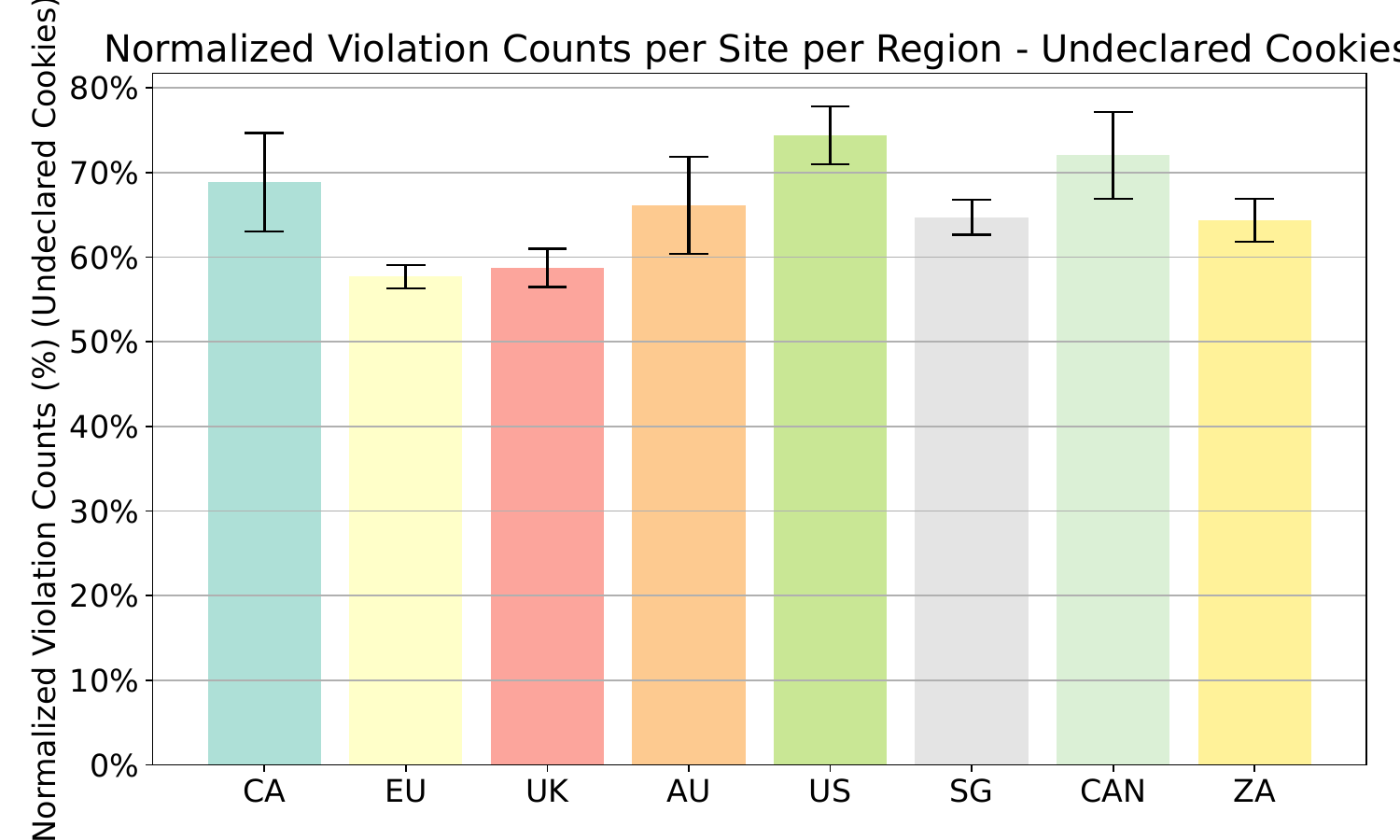}
        \caption{Undeclared Cookies}
        \label{fig:3rd_party_undeclared_cookie}
    \end{subfigure}
    \begin{subfigure}[t]{0.33\textwidth}
        \centering
        \includegraphics[width=\textwidth]{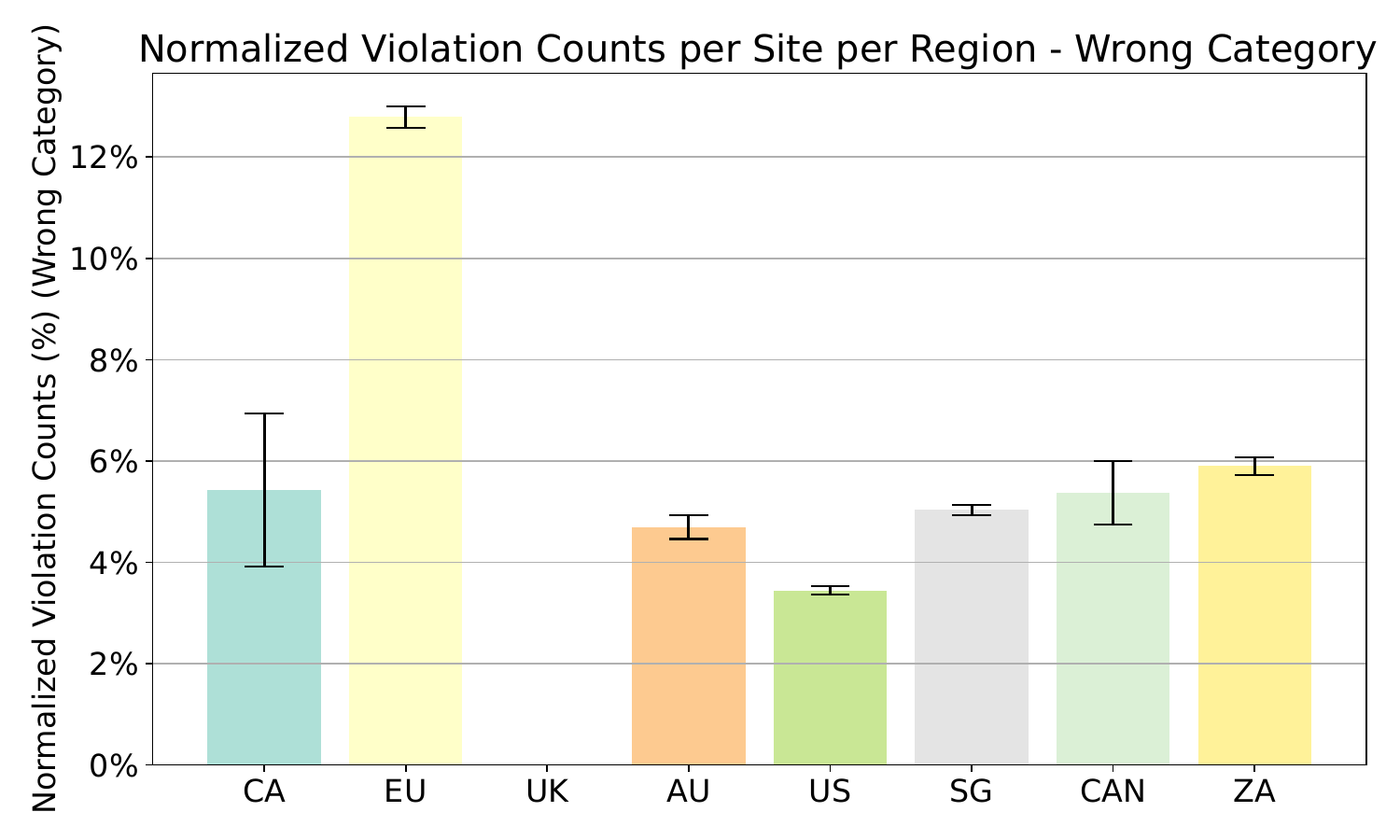}
        \caption{Wrong Cookie Category}
        \label{fig:3rd_party_wrong_category}
    \end{subfigure}
    \caption{Normalized average per-website 3rd-party cookie violations.}
    \label{fig:3rd_party_violations}
\vspace*{-0.07in}
\end{figure*}

Third-party cookies were most prevalent in the US across 
websites by a large margin (29.03 cookies per site). 
Regions with websites containing the fewest set of third-party cookies 
with only an average of 8.88 (EU) and 9.16 (UK) cookies per site. 
Other regions like CAN (22.87), AU (21.3), CA (20.26), SG (19.86), and ZA (19.52) had fewer third-party cookies 
loaded per website, but still significantly more than websites accessed from the 
EU and UK. These discrepancies may be attributed to a combination of 
advertisers' interest in the regions' users and region-specific privacy 
laws. CMPs provide website developers with templates and location rulesets to customize cookie 
banners and cookie placement depending on the region. Advertisers may 
have a higher interest in CA users for these English-speaking sites 
than regions in Asia or Africa (so cookies are more prevalent). Combined with CA's stricter 
requirements on opt-outs, personal information, and cookie notices, 
website developers may declare more cookies for the CA region, 
but place fewer on these users' browsers. 
North American regions have the most undeclared cookies, with CA 
containing the fewest. Whereas AU, SG, and ZA contain more 
undeclared cookies than the EU and UK. An illustrative example of this
is sportinglife.com which uses 3rd-party cookies from Google and 
DoubleClick. In CA, the website loads these undeclared cookies 
(e.g., NID, IDE) from both regions, whereas in the EU, 
the website does not load these cookies. 
Generally, the disparities observed in cookie consent violations 
(\Cref{fig:normalized_cookie_violations}) are exacerbated in 
3rd-party cookie consent violations (\Cref{fig:3rd_party_violations}).

\subsection{Cookie Banner 
and UI Discrepancies}

\begin{figure}[t]
    \centering
    \includegraphics[width=0.9\columnwidth]{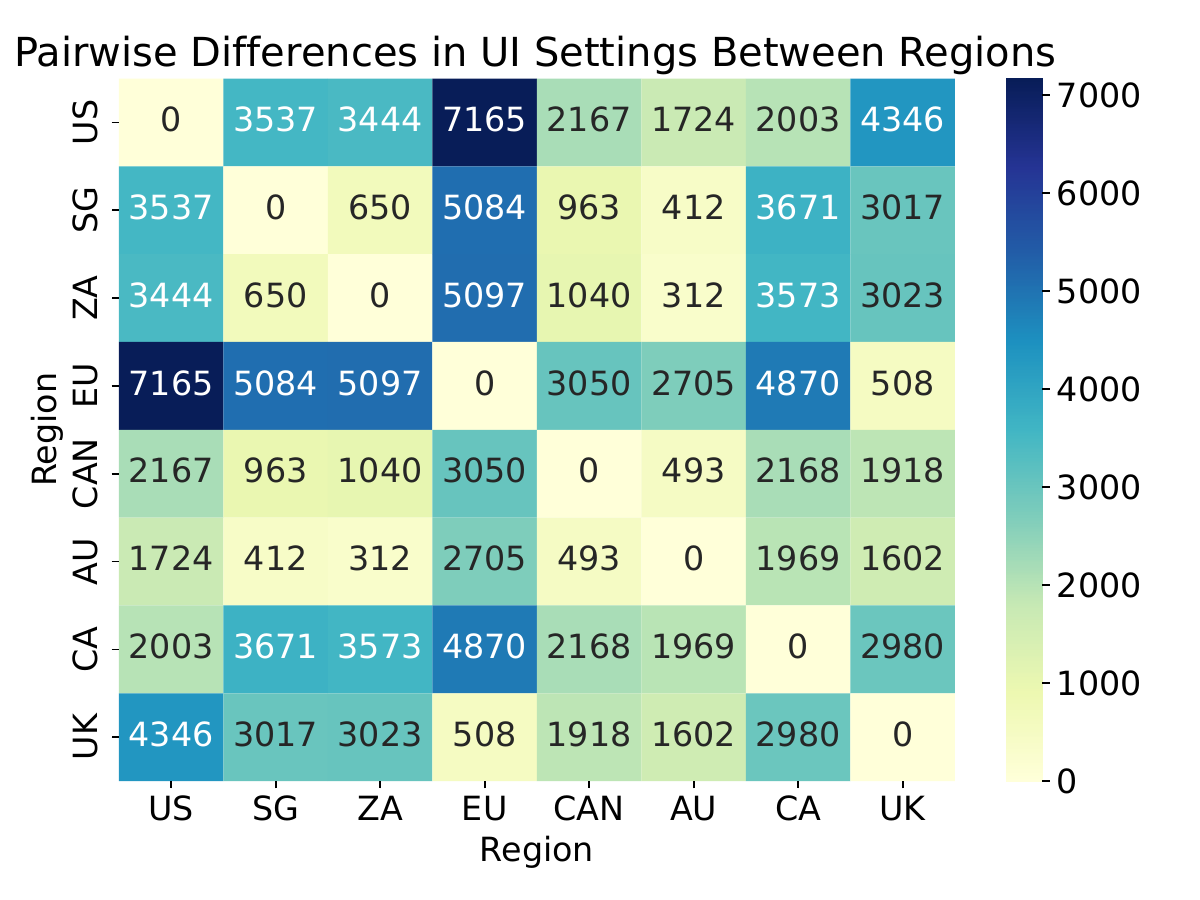}
    \vspace*{-0.08in}
    \caption{Pairwise website-level cookie banner implementation and UI differences across regions.}
    \label{fig:cookie_banner_ui_pairwise}
     \vspace*{-0.08in}
\end{figure}

We analyzed the OneTrust and Cookiebot cookie library 
configurations, both their consent choice and cookie banner UI settings. 
These banners were found to change a significant number of parameters 
such as cookie banner text, button color and positioning, banner 
background color and positioning, cookie consent model (opt-in 
vs.~opt-out, display banner vs.~implied consent choice, etc.), 
cookie category names, cookie lifespan, consent choice lifespan, 
and more. The EU and US (Non-CA) have the most disparities in 
cookie banner implementations (7,165), likely due to the lack of 
any privacy regulations in Michigan. SG (5084), ZA (5097), and CA (4870) also contain a significant number of cookie 
banner implementation differences, while Canada (3,050) and Australia (2,705) contain 
fewer differences. Larger differences emerge between the US and CA compared to all other 
regions likely stem from its importance as a region for 
advertisers, whereas the EU and UK's discrepancies likely stem 
from their privacy regulations and cookie guidelines. 
SG, ZA, CAN, and AU all have few pairwise differences relative 
to each other, ranging from 312 to 1040, far fewer than the US, 
CA, UK, and EU.

Many of these differences are minor changes in the text or UI colors, nudging users to select certain choices. For example, sites that display an ``ACCEPT ALL'' button on their cookie banner are much more likely to be accessed from the EU and UK, whereas in the US, ``OK'' is more commonly used. Other differences completely change the consent model of the cookie banner (e.g., ``opt-in'' vs. ``opt-out'' or ``1 month'' vs ``12 months'' for consent duration). In regions like the EU and UK, the probability of seeing a ``Reject All'' button on the cookie banner is higher than in other regions. Regions like CAN and SG, despite having explicit guidelines on cookie consent for online behavioral advertising, have a smaller share of sites' cookie banners using ``Reject All'' buttons.

Figures.~\ref{fig:cookie_banner_ui_pairwise} and 
10--16 show 
such differences in websites' cookie banner implementations. For 
example, Figures.~\ref{fig:ui_reject_all}, \ref{fig:ui_consent_lifetime}, and
\ref{fig:ui_consent_model} 
all indicate that websites in the EU and UK tend to use more 
privacy-preserving practices and features, such as the consent 
choice lasting longer, giving users the option to 
reject all cookies, having both 1st-party and 3rd-party cookies 
be opt-in rather than opt-out or an implied consent choice.


\section{Root Cause Analysis}
\label{sec:root_cause}

We analyzed Onetrust and Cookiebot's documentation and demos to gain 
insight into whether CMPs or website developers hold 
responsibility for cookie consent violations. Although CMPs may act as 
data controllers or processors~\cite{santos2021consent}, both parties 
share responsibility. CMPs need to provide stronger guardrails, easier 
integration methods, support for more legal frameworks, and clearer 
documentation. Site owners need to ensure no undeclared 
cookies remain, properly implement cookie/script-blocking, and keep 
their CMP integrations updated. 
We find that most undeclared cookies result from either website 
developers not categorizing cookies or CMP crawlers missing some 
cookies, whereas ignored cookie rejections stem from 
developers not integrating CMP scripts correctly.

\subsection{CMP Responsibility}
\noindent
{\em\underline{\bf Template Gaps}}:
Most CMPs ship with templates only for major legal frameworks 
like GDPR and CCPA, leaving other jurisdictions unsupported. 
As a result, developers have no pre-configurations or guidelines 
for regions like Canada, Singapore, or South Africa, leading to 
inconsistent or noncompliant cookie banners 
(\Cref{fig:cmp_templates}). 
Note that these templates are useful for developers (unfamiliar with 
privacy laws) in determining cookie banner behaviors (e.g., whether they are shown, reconsent expiration time, etc. \Cref{fig:cmp_rule_behaviors}).

\noindent
{\em\underline{\bf Incomplete Automatic Crawling}}:
CMPs commonly auto‐detect 3rd‐party scripts/iframes to populate 
banner catalogs and block rejected categories. 
However, content loaded only in specific contexts (e.g., embedded 
YouTube videos) can evade these crawlers, so cookies can slip through undetected, leading to undeclared cookie violations
(Table~\ref{tab:top_incor_receivers}).

\noindent
{\em\underline{\bf Geolocation Rulesets}}:
CMPs offer geolocation rulesets that dynamically adjust banner text, 
button labels, consent models (opt-in vs.\ implied), and even hide 
“Reject All” to maximize opt-in rates per region 
(\Cref{fig:cmp_geolocation_rules}). 
CMPs support geolocation rulesets, which can allow for customizing 
cookie banner stylization and functionality depending on the region 
of the user's IP address. 
For example, the CCPA template for Onetrust supports translating 
behaviors like clicking or moving to the next page as the consent 
choice "Accept All Cookies" 
(\Cref{fig:cmp_rule_behaviors}). 
At other locations without legal template support, website developers
may set geolocation rulesets to not display cookie banners, 
or even allow scrolling to be counted as accepting all cookies. 
These rulesets account for the disparity in cookie placement 
and consent functionality between regions.

\noindent
{\em\underline{\bf Dark Patterns and Opt-In Analytics}}:
Cookiebot allows web developers to change cookie banner text and 
cookie banner stylization without guardrails. 
Developers can set arbitrary labels
for buttons, for example, some sites 
used "Continue to Site" and "Close" as labels in place of 
"Accept All Cookies". Further, CMP dashboards provide A/B testing 
and opt-in analytics, incentivizing developers to create dark 
patterns and optimizing banner layouts for acceptance rate.

\subsection{Developer Responsibility}

\noindent
{\em\underline{\bf Incorrect Script Embedding}}:
In order for CMPs to block 3rd-party scripts that load cookies, 
website developers need to update any <script> tags with additional 
attributes (e.g., data-usercentrics="Google Maps" or <script 
class="optanon-category-C0002">).
Additionally, CMP scripts connected to the Onetrust and Cookiebot 
services must be placed in the <head> \textit{before} any other 
scripts~\cite{ashlea_cartee_say_2019,cybot_automatic_2019,cookiepro_onetrust_2021}. If these instructions are not followed 
perfectly, scripts will continue to load and cookies will be placed. 
Both OneTrust and Cookiebot's documentation states that if cookies 
or tracking scripts are loaded before consent withdrawal,
then the developer or user needs to deactivate the service or 
perform a page reload~\cite{onetrust2024docs,cookiebot2023docs}.
A manual inspection revealed that websites left scripts uncontrolled 
by the CMPs. For example, on \textit{scientificamerican.com}, 
we rejected \textit{Performance cookies}, but found that both 
first- and third-party Analytics cookies such as Google Analytics 
(\textit{\_ga}) were still being loaded 
(Table~\ref{tab:top_incor_cookies}).


\noindent
{\em\underline{\bf Manual Cookie Categorization}}:
When CMP auto-scanners miss new or previously uncategorized cookies, 
developers need to manually assign them to categories. 
In practice, these manual steps could be easily missed or skipped, 
so these “unknown” cookies remain undeclared and active. 
These missed cookies likely make up the bulk of the undeclared
cookie violations (\Cref{fig:cmp_manual_defs}).

\noindent
{\em\underline{\bf Neglecting Updates}}:
Websites frequently change~\cite{adar2009web}, and web developers may forget to have CMPs 
re-crawl their sites and update their scripts, especially on new subdomains. 
The Cookiebot scanner is also priced to scale with the number of 
subpages to crawl, and thus may provide incomplete scans. 
These may also contribute to undeclared cookies.

\noindent
{\em\underline{\bf Banner Localization Oversights}}:
Beyond CMP geolocation tweaks, developers may hardcode or 
override banner text, styling, and behaviors for each locale.

\section{Conclusion}

We presented \sys, an automated system that 
detects cookie consent violations on websites across the world.
We also developed a formal model to systematically
analyze the (in)consistencies between users' cookie consent 
preferences and actual cookie usage of websites.
Our findings indicate that the majority of the studied 
websites contain inconsistent and potentially non-compliant
cookie consent behavior in the measurements from 8 regions.
Our findings suggest the existence of systemic issues with 
cookie banners and CMPs, highlighting the need for 
larger-scale auditing and enforcement of cookie usage 
to protect users' autonomy and privacy. 

\section{Ethics Considerations}

Our study was conducted with strict adherence to ethical research practices, ensuring minimal impact on the websites and their users. All data was obtained from publicly accessible websites via browser automation tools (e.g., Playwright, Puppeteer). We crawled the sites and performed our measurements without storing their HTML or page data. The study focused solely on analyzing CMP libraries and cookie data.

To conduct a comprehensive analysis across regions, we utilized AWS and DigitalOcean proxies to access websites from different geographic locations. Basic bot detection evasion techniques were employed to increase coverage. Our crawling processes were designed to generate minimal traffic, only repeating the measurements 10 times to ensure reproducibility. We avoided straining servers and disrupting normal website operations by spacing the measurements over a week.

\section{Compliance with Open Science Policy}

To uphold principles of transparency and reproducibility, the source code of our crawling and measurement system, alongside the analysis scripts and cookie data are available at \url{https://github.com/byron123t/cookie-consent} or \url{https://doi.org/10.5281/zenodo.15566975}.

However, certain components of our system are subject to patents and institutional licensing restrictions, which may prevent full functionality of the shared code. We will make the code functional without these components, with this code being available upon request and approval by our institution.

\section{Acknowledgements}
The work reported in this paper was supported in part by the
National Science Foundation under Grant CNS-2245223.


\printbibliography
\pagebreak
\appendix

\begin{table*}[t]
\centering
\footnotesize
\centering
\begin{tabular}{lrrrrrrrr}
\toprule
\% Cookies & CA & EU & UK & AU & US & SG & CAN & ZA \\
\midrule
Trackers & 63.87\% & 64.10\% & 64.14\% & 63.36\% & 64.73\% & 62.89\% & 64.11\% & 64.40\% \\
Location & 7.34\% & 7.44\% & 7.38\% & 7.93\% & 8.18\% & 7.99\% & 8.19\% & 7.46\% \\
IP Address & 3.28\% & 5.74\% & 5.58\% & 3.51\% & 2.81\% & 3.54\% & 3.01\% & 3.81\% \\
Language & 0.54\% & 1.00\% & 0.92\% & 0.52\% & 0.45\% & 0.55\% & 0.51\% & 0.57\% \\
Unlikely P.I. & 24.98\% & 21.98\% & 21.30\% & 23.84\% & 25.03\% & 25.50\% & 24.09\% & 24.76\% \\
\bottomrule
\end{tabular}
\caption{Violations and cookies containing personal information in the form of trackers, ip, or location data.}
\label{tab:personal_information}
\vspace{-.2cm}
\end{table*}


\begin{table}[t]
    \centering
\begin{minipage}[b]{0.45\linewidth}
    \footnotesize
    \centering
    \begin{tabular}{lr}
    \toprule
    Cookie Name &  \# Websites (\%) \\
    \midrule
            \_ga &    583 (48.4) \\
           \_gid &    429 (35.6) \\
           \_fbp &    318 (26.4) \\
            IDE &     296 (24.6) \\
       \_uetsid &     218 (18.1) \\
    \bottomrule
    \end{tabular}
    \caption{Top-5 Rejected Cookies Used.}
    \label{tab:top_incor_cookies}
\end{minipage}
\hfill
\begin{minipage}[b]{0.5\linewidth}
    \footnotesize
    \centering
    \begin{tabular}{ll}
    \toprule
            Tracker &  \# Websites (\%) \\
    \midrule
    doubleclick.net & 328 (27.2) \\
       linkedin.com & 198 (16.4) \\
        youtube.com & 198 (16.4) \\
           bing.com &   88 (7.3) \\
        twitter.com &   69 (5.7) \\
    \bottomrule
    \end{tabular}
    \caption{Top trackers of rejected-usage cookies.
    }
    \label{tab:top_incor_receivers}
\end{minipage}
\end{table}

\remove{
\begin{table}[t]
    \small
    \centering
\begin{tabular}{lrr}
\toprule
Cookie Duration &  Frequency (\%) &  \% of ``Too-Short Cookies'' \\
\midrule
        Session &          23.27 &                             8.02 \\
  A few seconds &          22.58 &                            89.91 \\
       365 days &          11.75 &                             5.83 \\
       730 days &           4.55 &                             3.81 \\
        30 days &           3.97 &                            10.49 \\
\bottomrule
\end{tabular}
    \caption{
    Most common durations in cookie declarations.}
    \label{tab:common_cookie_durations}
\end{table}
}



\begin{figure}[t]
    \centering
    \includegraphics[width=0.8\columnwidth]{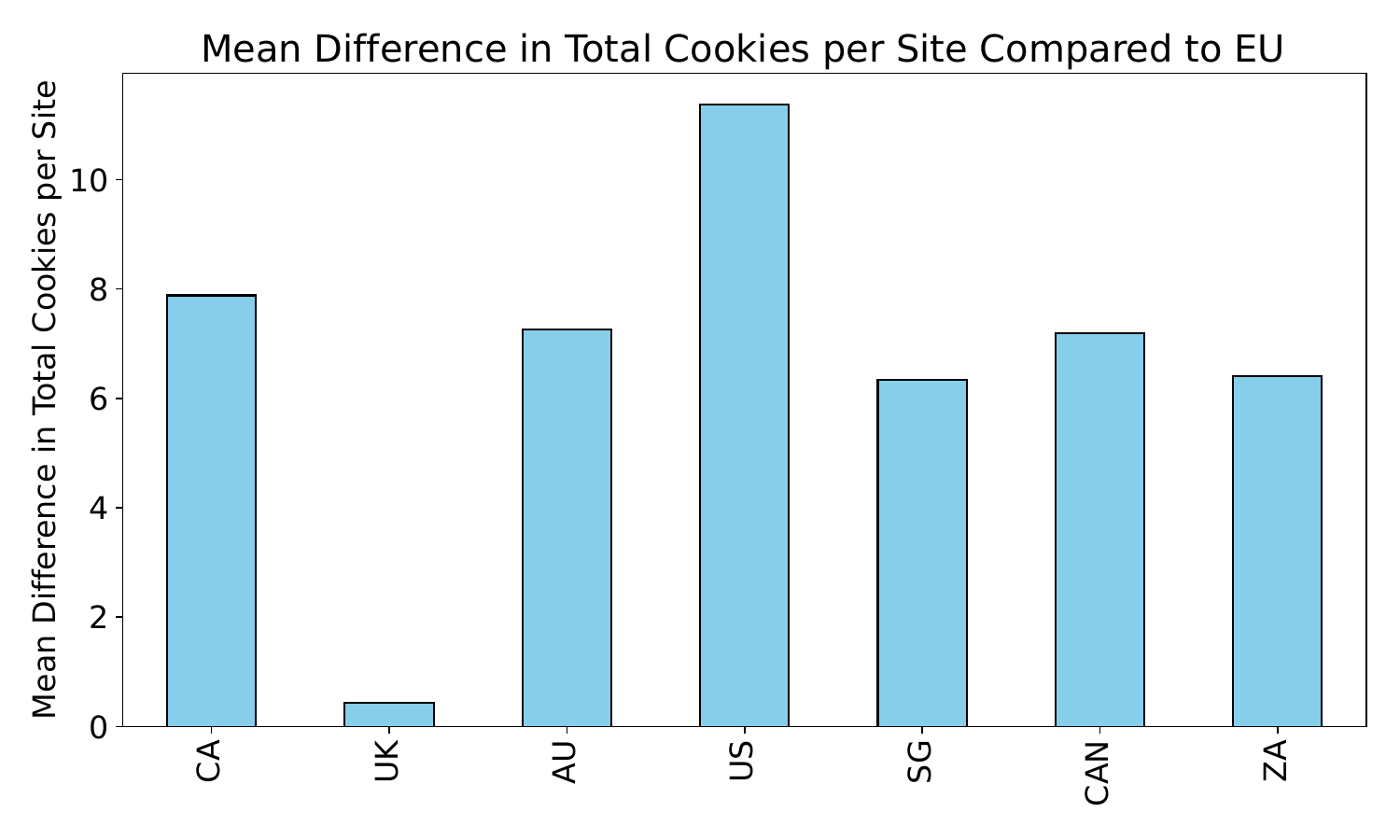}
    \caption{Relative website-level differences in total cookies. (Compared to EU baseline)}
    \label{fig:difference_cookies}
\vspace*{-0.07in}
\end{figure}
\begin{figure}[t]
    \centering
    \includegraphics[width=0.8\columnwidth]{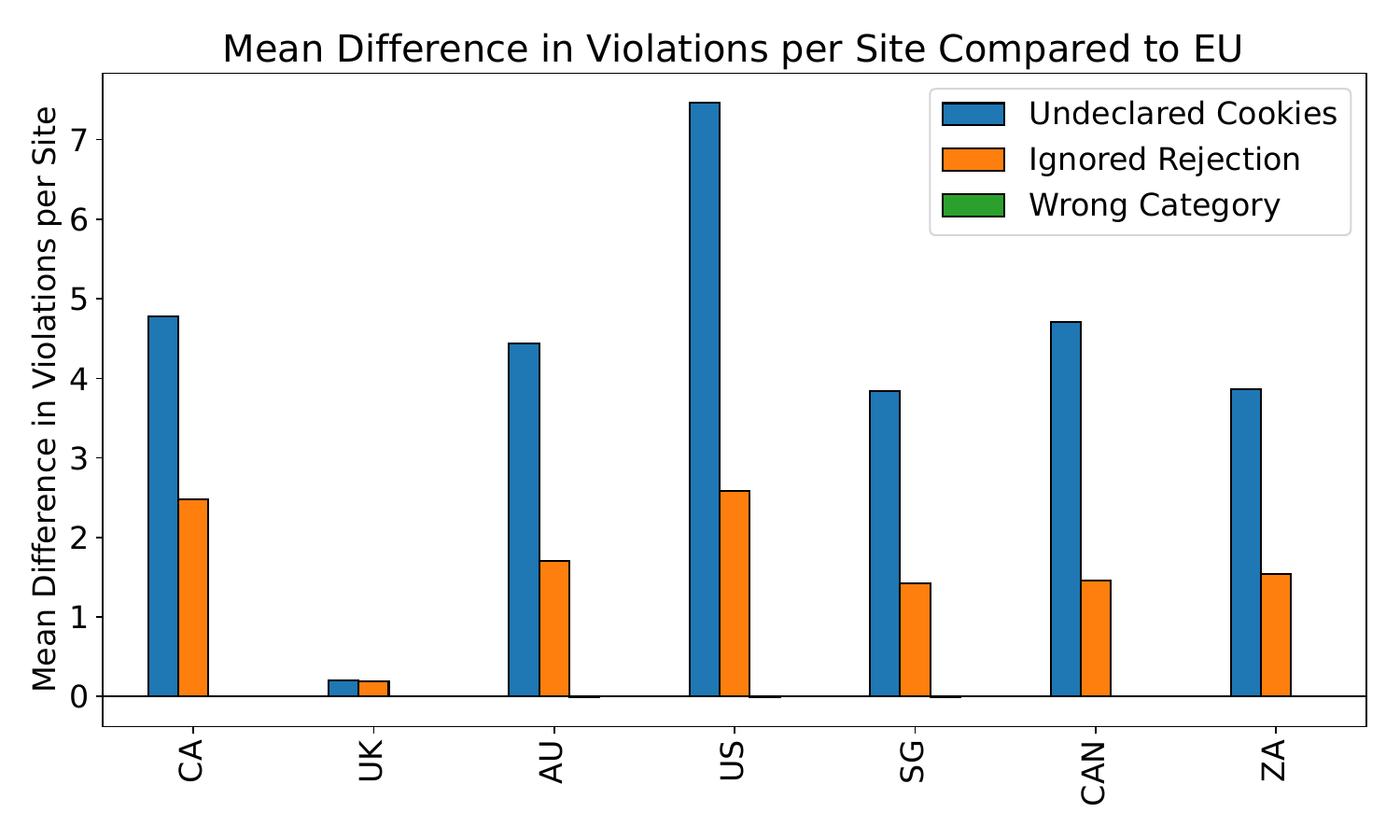}
    \caption{Relative website-level differences in cookie violations. (Compared to EU baseline)}
    \label{fig:difference_cookie_violations}
\vspace*{-0.07in}
\end{figure}


\begin{figure}[t]
    \centering
    \includegraphics[width=\columnwidth]{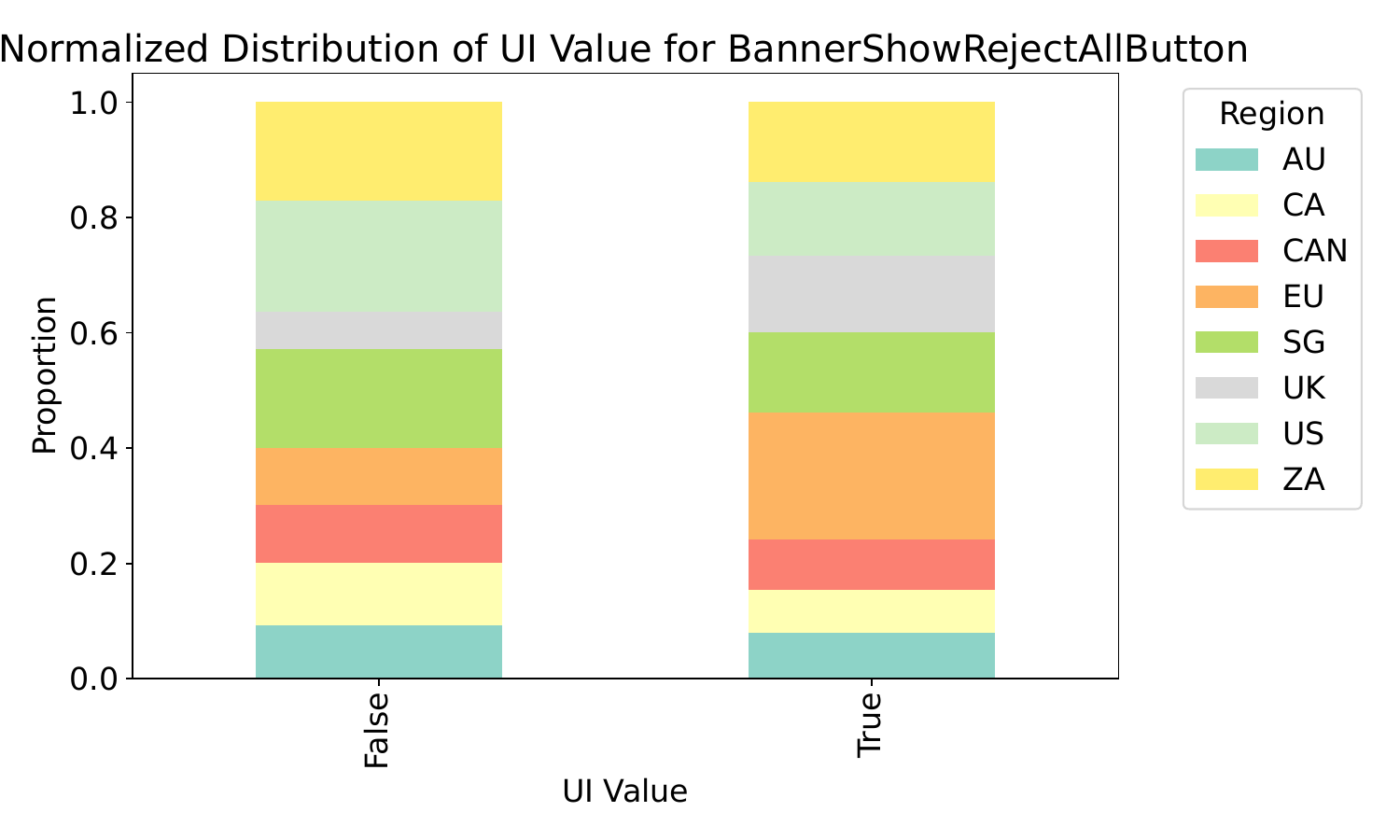}
    \caption{Reject all button presence in cookie banners.}
    \label{fig:ui_reject_all}
\end{figure}

\begin{figure}[t]
    \centering
    \includegraphics[width=\columnwidth]{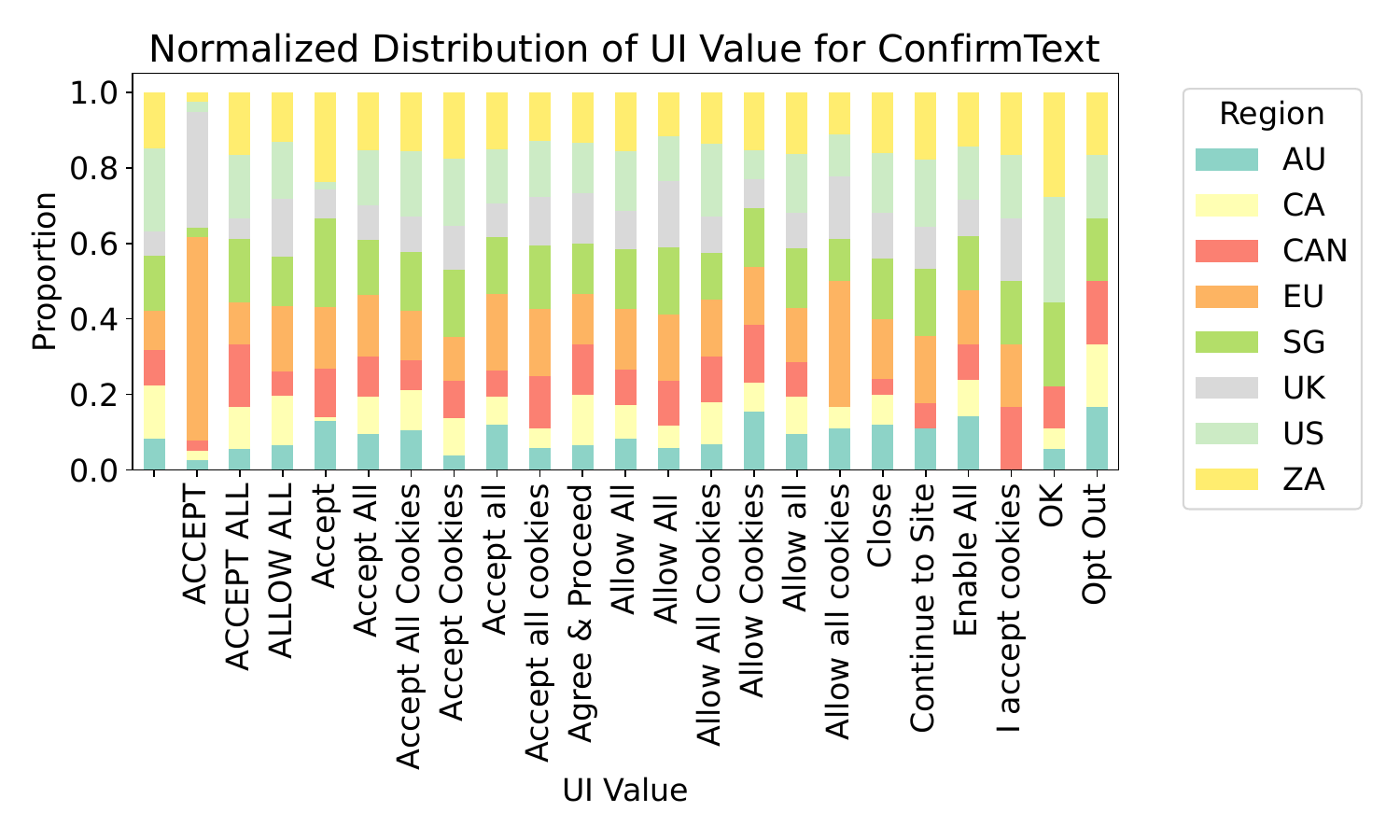}
    \caption{Pairwise website-website cookie count differences.}
    \label{fig:ui_confirm_text}
\end{figure}

\begin{figure}[t]
    \centering
    \includegraphics[width=\columnwidth]{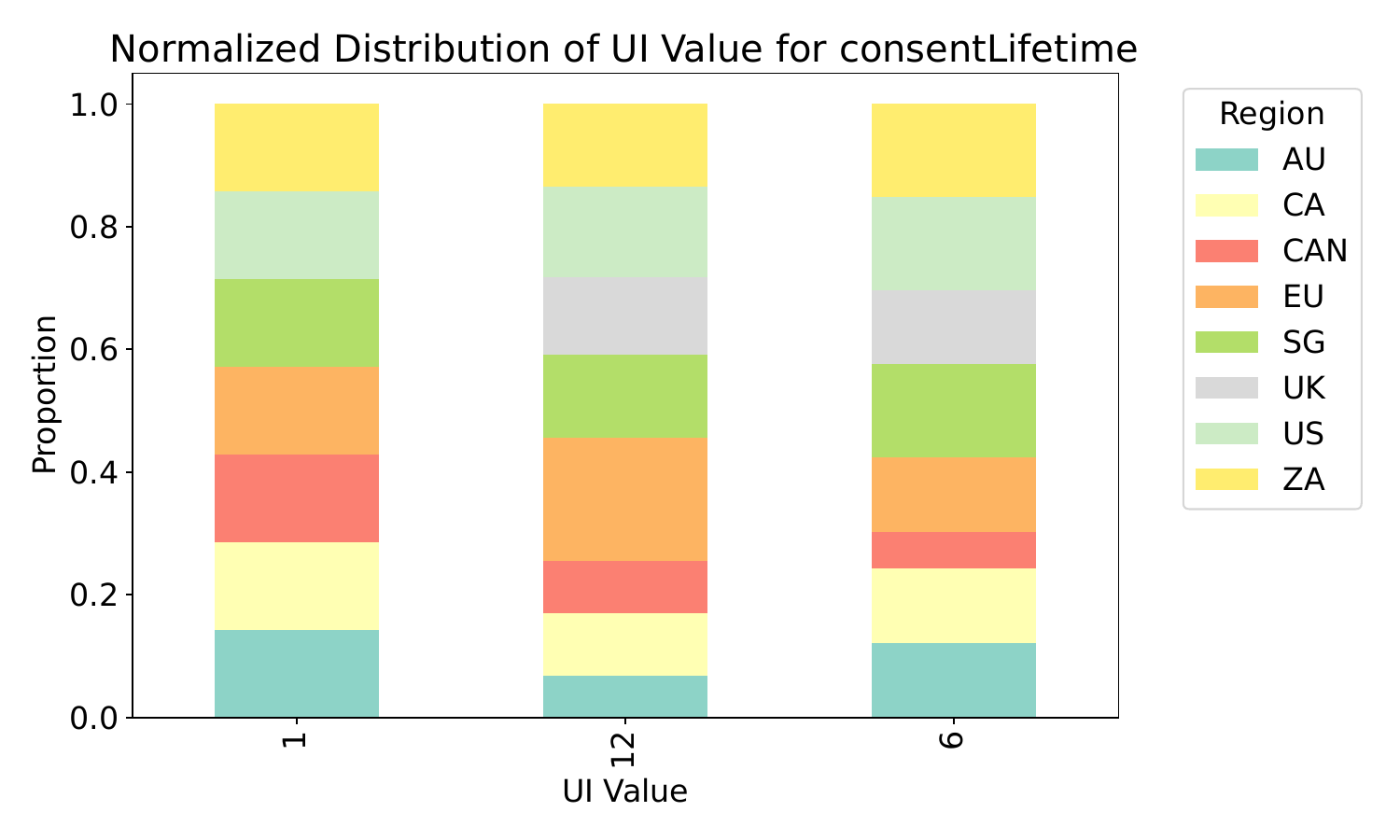}
    \caption{Pairwise cookie consent lifetime differences.}
    \label{fig:ui_consent_lifetime}
\end{figure}

\begin{figure}[t]
    \centering
    \includegraphics[width=\columnwidth]{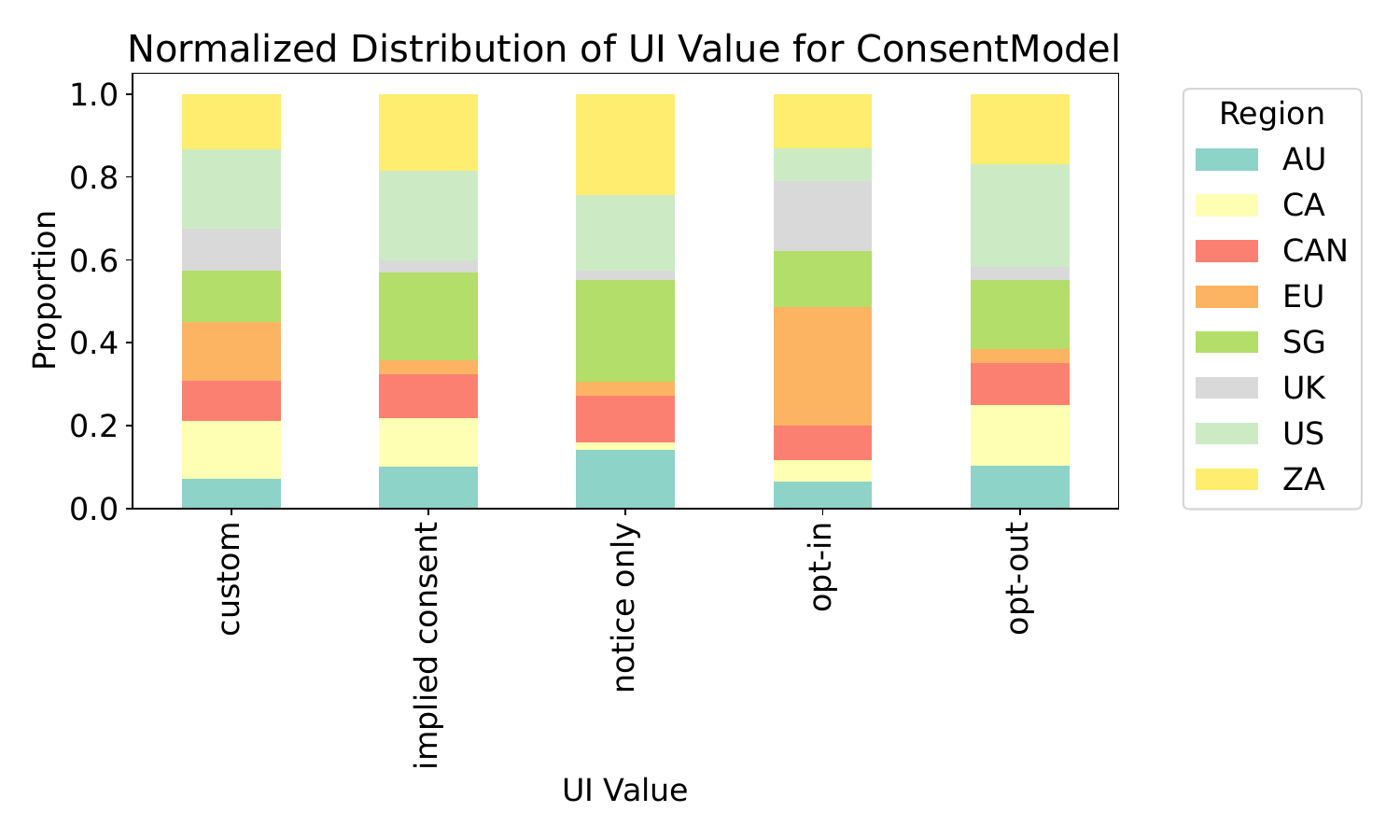}
    \caption{Pairwise CMP behavior differences.}
    \label{fig:ui_consent_model}
\end{figure}




\begin{figure*}[t]
    \begin{minipage}[b]{0.25\linewidth}
    \centering
    \includegraphics[width=\linewidth]{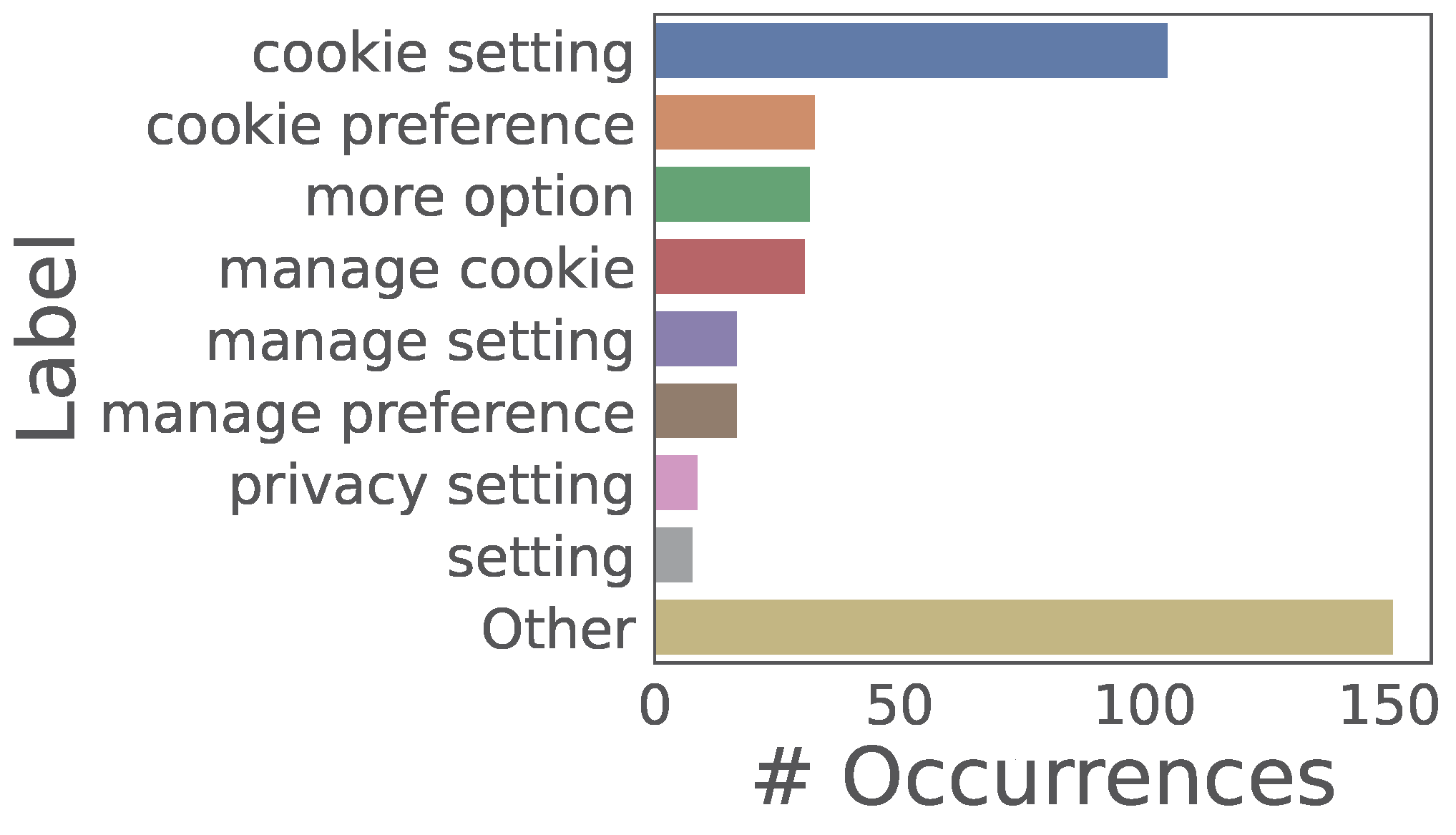}
    \caption{Distribution of labels of cookie banner buttons.}
    \label{fig:preference_button_label_dist}
    \end{minipage}
    \hfill
    \begin{minipage}[b]{0.42\linewidth}
    \scriptsize
    \centering
    \begin{tabular}{lrrr}
    \toprule
    G & Feature (Dimension) &  HTML attributes & $D_G$ \\
    \midrule
    $G_1$ &  \# n-grams and keywords (3) & aria-label, class, id, text & 12 \\
    $G_2$ &  \# tokens $> n_t$ or not (1) & aria-label, text  & 2 \\
    $G_3$ &  Has consent library API (1) & class, href, id, onclick & 3 \\
    \midrule
      & & Total & 17 \\
    \bottomrule
    \end{tabular}
    \captionof{table}{Cookie button detection features.
    \textit{G} and $D_G$ stand for 
    a feature group and its dimension.
    }
    \label{tab:pref_btn_detect_features}
    \end{minipage}
    \hfill
    \begin{minipage}[b]{0.28\linewidth}
    \centering
    \includegraphics[width=\linewidth]{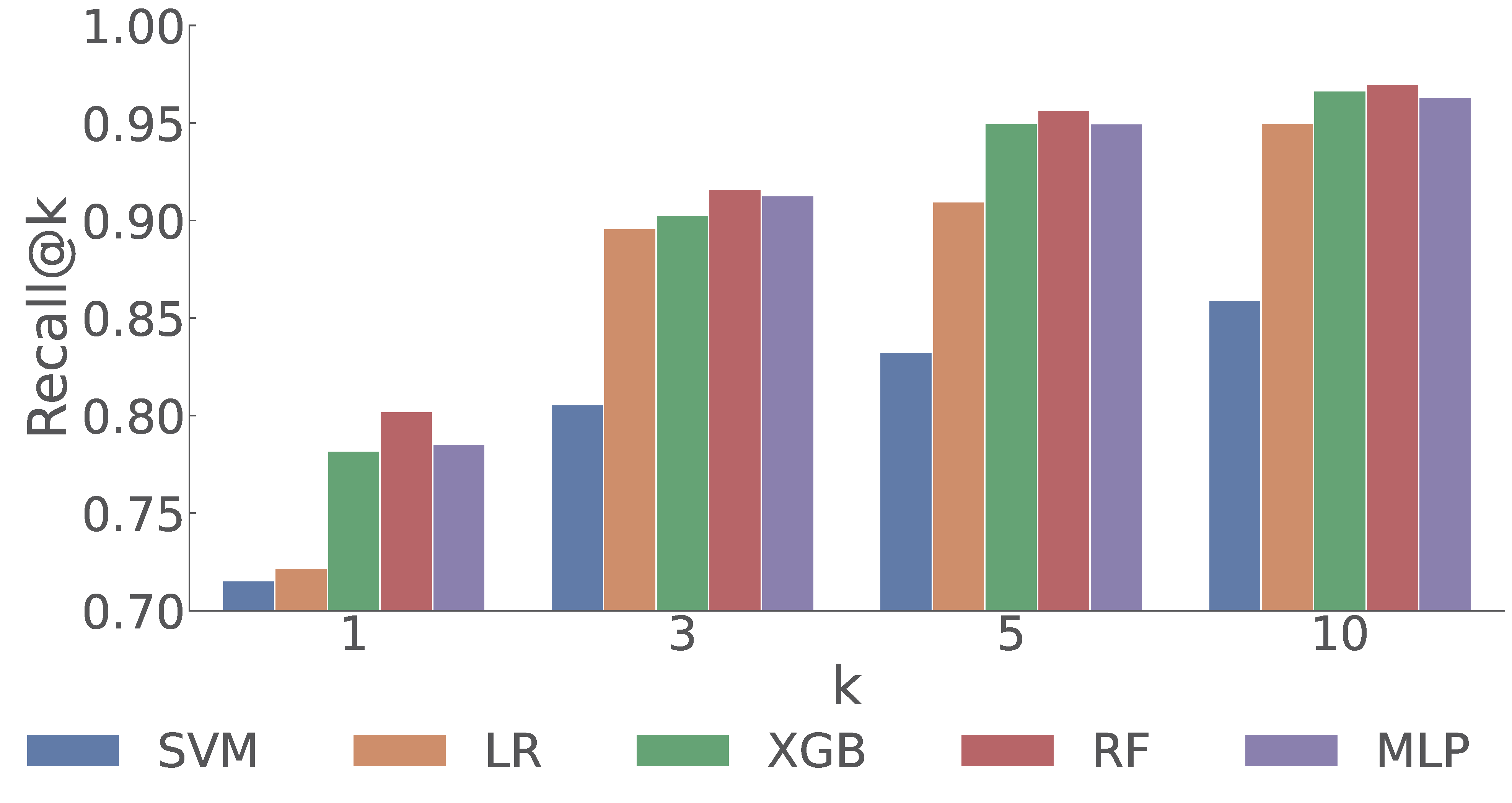}
    \caption{Top-k scores of 10-fold validation of ML models.}
    \label{fig:model_selection_topk}
    \end{minipage}
\end{figure*}

\section{Preference Button Extractor}
\label{sec:pref_btn}

\subsection{Button Extractor Data Collection}
\label{sec:pref_btn_data_coll}

\begin{table}[tbh]
    \footnotesize
    \centering
    \begin{tabular}{p{.1\linewidth}p{.8\linewidth}}
    \toprule
    Group & Examples \\
    \midrule
    Unigrams & adchoice, adjust, change, choice, choose, configure, consent, cookie, customise, customize, manage, option, personal, preference, privacy, review, setting, update, view \\
    Bigrams & configure consent, set preference, advanced setting, privacy setting, update preference, personal information, manage preference, california sell, privacy preference, sell personal, consent detail, manage setting, change privacy, view cookie \\
    Keywords & change consent, change setting, consent choice, consent tool, cookie consent, cookie preference, cookie setting, customize setting, manage cookie, review cookie \\
    \bottomrule
    \end{tabular}
    \caption{Examples of n-grams and high-frequency keywords
        extracted from the button labels.
    } \label{tab:pref_btn_detect_keywords}
\end{table}

We randomly selected 8k websites from the top 
\numSitesAvailStudy global websites for developing 
features and training the button extractor. The remaining 2k 
websites were set aside to evaluate the model performance. 
We use the Tranco list~\cite{LePochat_tranco_2019} 
generated in July 2021 (ID: 9QK2) and accessed websites 
from an IP address in the UK to maximize encounters with 
cookie banners and preference buttons. We manually visited the home pages of 
\lyx{1,000} randomly selected websites from the 8k 
websites to identify cookie banner buttons. Two 
annotators independently annotated the home pages 
of the websites. For each website, a snapshot of the 
home page HTML and the CSS selectors were recorded. Only 
the websites with English home pages were annotated, 
excluding non-English pages and duplicates. We obtained a 
training set of \lyx{298} web pages containing \lyx{436} 
cookie banner buttons out of \lyx{71,020} all links/buttons. 
Many websites only show a cookie banner without any choice 
or only a binary accept/reject option.


\subsection{Button Extractor Feature Selection}
\label{sec:pref_btn_feat_sel}

We derive \lyx{3} classification feature groups based on 
the HTML attributes: aria-label, class, id, and inner 
text. The attribute \textit{aria-label}, an accessibility 
feature of the web for marking buttons with labels for 
users with disabilities~\cite{cookiebot_functions_2021}, 
is especially useful in cases where buttons are displayed 
as non-textual icons. The features are shown in 
Table~\ref{tab:pref_btn_detect_features} and the feature 
vector has \lyx{17} dimensions in total. Feature group 
$G_1$ is the number of occurrences of selected unigrams, 
bigrams and high-frequency keywords in the button labels. 
We also separate the most frequently-used bigrams into a 
set of high-frequency keywords 
(Table~\ref{tab:pref_btn_detect_keywords} lists some 
examples). Feature group $G_2$ indicates whether or not 
the number of tokens of a button label is greater than a 
threshold $n_t=9$ (empircally selected using training 
set). This feature allows our classifier to avoid long 
paragraphs that may contain keywords. Finally, feature 
group $G_3$ indicates whether a cookie consent library API 
is being used.

\subsection{Button Extractor Performance Metric}
\label{sec:pref_btn_perf_metric}

We report the \lyx{\textit{recall@k}} score, a metric 
used to evaluate information retrieval systems
\cite{polisis_usenix_2018,thung2013automated,sun2011towards}, 
which represents the portion of websites containing 
cookie banner buttons detected from the top \textit{k} 
classification results. A website is successfully detected 
if one of its cookie banner buttons is among the top 
\textit{k} buttons with the highest classification 
probabilities.

\subsection{Button Extractor Model Selection}
\label{sec:pref_btn_model}

After evaluating various ML algorithms including 
regression, perceptrons, SVMs, etc., we find that the 
best-performing classifier is a random forest (RF) with 
100 decision trees which achieves the best 
\textit{recall@1} score of \lyx{80.22\%} ($1 \le k \le 
10$, 10-fold cross-validation). It consistently 
outperformed other models \textit{recall@k} for $k \in \{1,3,5,10\}$. Fig.~\ref{fig:model_selection_topk} 
shows the \textit{recall@k} scores of the models. 
Using \lyx{57} websites containing preference buttons 
from the \lyx{2k} domains in the test set, we find that 
the performance of the model achieves \textit{recall@1}, 
\textit{recall@3}, \textit{recall@5} and 
\textit{recall@10} scores of \lyx{77.19\%}, 
\lyx{\topThreeScore}, \lyx{85.96\%}, and \lyx{89.47\%}. These are the recall scores for the top 3, 5, and 10 detected objects.

\section{Consent Cookie Decoding}
\label{appd:decode_consent_resources}

To extract cookie consent preferences,
we decode consent cookies basing on the documentation and analyzing their key-value pairs.
\onetrust's consent cookie is called \textit{OptanonConsent}~\cite{onetrust_llc_onetrust_2021}
which stores the consent preference
of each cookie category. For example, \textit{groups=C1:1,C2:0}
indicates that cookie category $C1$ is approved while $C2$ is rejected.
\cookiebot's consent cookie is called \textit{CookieConsent}
storing consents for 4 fixed cookie categories: Necessary, Preferences, Statistics, and 
Marketing~\cite{cybot_developer_2021}.
Similar to 'Necessary' cookies,
'Unclassified' cookies are not automatically blocked and cannot be denied by users, so the consent preferences for these cookies are by default set to True~\cite{cybot_cookiebot_2021}. 
\label{appendix:cmp_docs}

\begin{figure}[t]
    \centering
    \includegraphics[width=0.95\columnwidth]{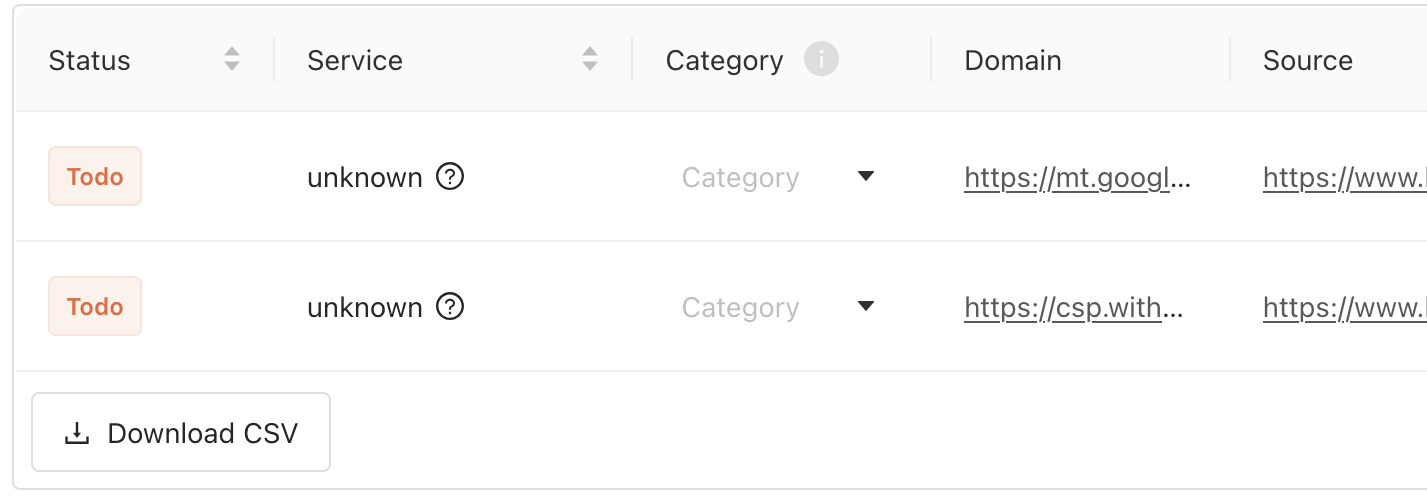}
    \caption{Cookiebot developer interface for handling uncategorized and undeclared cookies.}
    \label{fig:cmp_manual_defs}
\end{figure}

\begin{figure}[t]
    \centering
    \includegraphics[width=0.95\columnwidth]{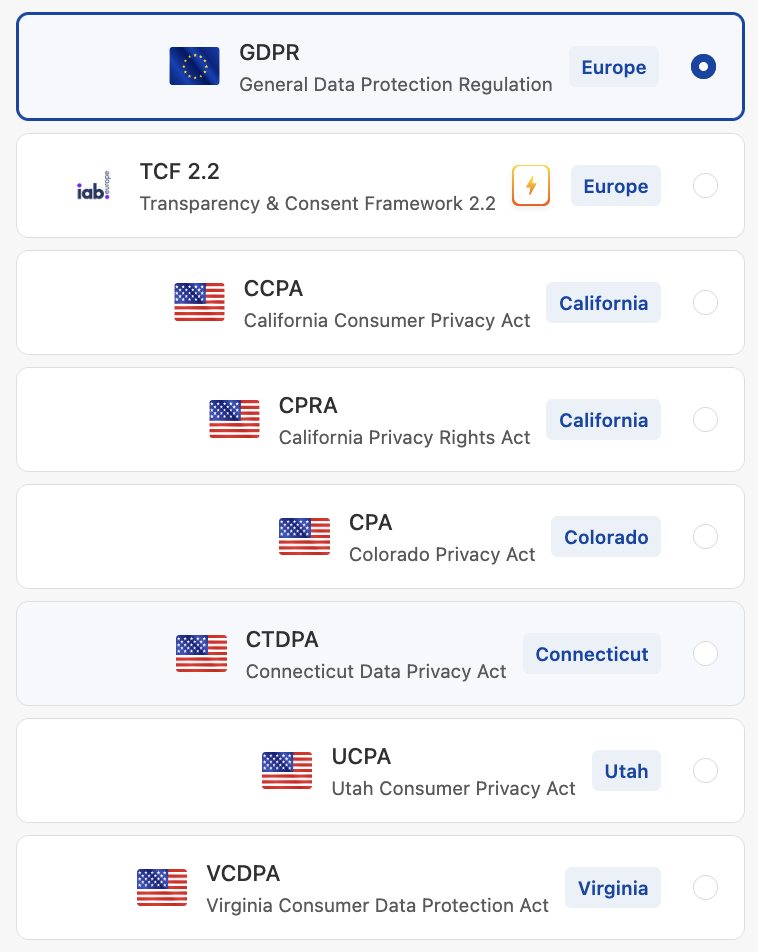}
    \caption{Cookiebot options for regional privacy law templates and cookie banner implementation guidelines. Missing: Singapore, Canada, Brazil, South Africa, etc.}
    \label{fig:cmp_templates}
\end{figure}

\begin{figure}[t]
    \centering
    \includegraphics[width=0.95\columnwidth]{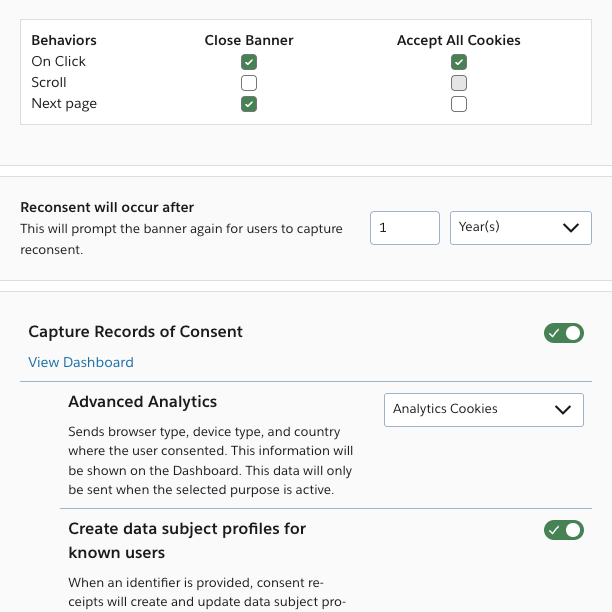}
    \caption{Onetrust developer interface for consent expiration and automatically setting consent on user behaviors.}
    \label{fig:cmp_rule_behaviors}
\end{figure}

\begin{figure}[t]
    \centering
    \includegraphics[width=0.95\columnwidth]{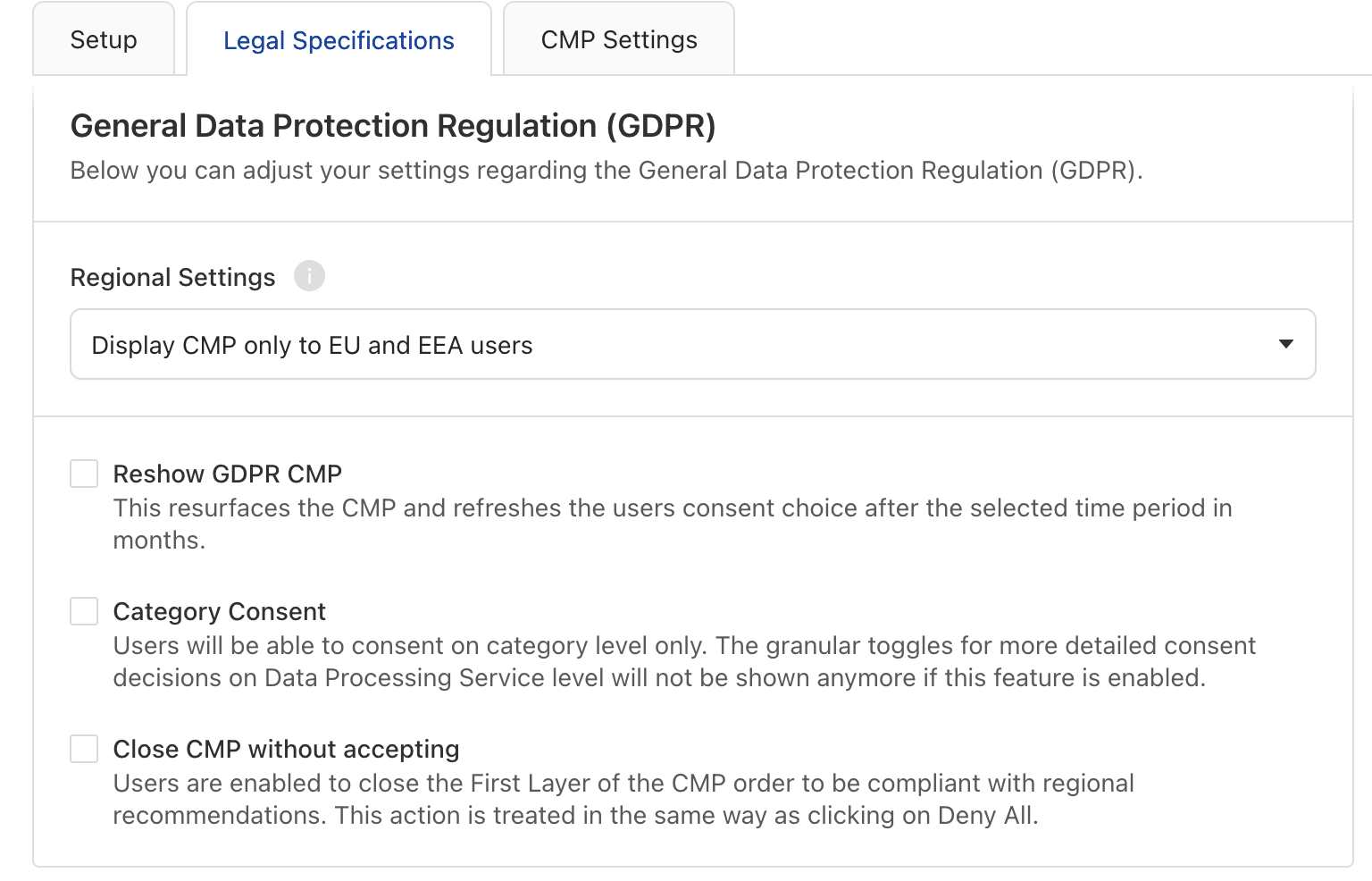}
    \caption{Cookiebot geolocation rulesets and default options for CMP display.}
    \label{fig:cmp_geolocation_rules}
\end{figure}





\end{document}